\documentclass[twocolumn]{jpsj3} 
\usepackage{graphicx}
\usepackage{bm}

\title{High-$T_c$ Superconductivity and Antiferromagnetism in Multilayered Copper Oxides\\
 $-$ A New Paradigm of Superconducting Mechanism $-$ }

\author{Hidekazu Mukuda\thanks{E-mail: mukuda@mp.es.osaka-u.ac.jp}, Sunao Shimizu\thanks{Present address: CERG, RIKEN, Wako 351-0198, E-mail: sshimizu@riken.jp}, Akira Iyo$^{1}$\thanks{E-mail: iyo-akira@aist.go.jp}, and Yoshio Kitaoka\thanks{E-mail: kitaoka@mp.es.osaka-u.ac.jp} }

\inst{Graduate School of Engineering Science, Osaka University, Toyonaka, Osaka 560-8531 \\
$^1$National Institute of Advanced Industrial Science and Technology (AIST), Umezono, Tsukuba 305-8568 \\
}

\abst{
High-temperature superconductivity (HTSC) in copper oxides emerges on a layered CuO$_2$ plane when an antiferromagnetic  Mott insulator is doped with mobile hole carriers. We review extensive studies of multilayered copper oxides by site-selective nuclear magnetic resonance (NMR), which have uncovered the intrinsic phase diagram of antiferromagnetism (AFM) and HTSC for a disorder-free CuO$_2$ plane with hole carriers. We present our experimental findings such as the existence of the AFM metallic state in doped Mott insulators, the uniformly mixed phase of AFM and HTSC, and the emergence of $d$-wave SC with a maximum $T_c$ just outside a critical carrier density, at which the AFM moment on a CuO$_2$ plane disappears. 
These results can be accounted for by the {\it Mott physics} based on the $t$-$J$ model. The superexchange interaction $J_{\rm in}$  among spins plays a vital role as a glue for Cooper pairs or mobile spin-singlet pairs, in contrast to the phonon-mediated attractive interaction among electrons established in the Bardeen-Cooper-Schrieffer (BCS) theory. We remark that the attractive interaction for raising the $T_c$ of HTSC up to temperatures as high as 160 K is the large $J_{\rm in}$ ($\sim 0.12$ eV), which binds electrons of opposite spins to be on neighboring sites, and that there are no bosonic glues. It is the Coulomb repulsive interaction $U$~($ > 6$ eV) among Cu-3$d$ electrons that plays a central role in the physics behind high-$T_{\rm c}$ phenomena. A new paradigm of the SC mechanism opens to strongly correlated electron matter.  
}

\kword{copper oxide, correlated electrons, high-$T_{\rm c}$ superconductivity, antiferromagnetism, phase diagram, NMR}

\begin{document}
\maketitle

\date{\today}

\section{Introduction}

In 1986, Bednorz and M\"uller unveiled a new class of ceramic superconducting materials composed of a layered structure of a two-dimensional CuO$_2$ plane\cite{BM}. 
Subsequently, the superconducting transition temperature ($T_{\rm c}$) in cuprates rises up to $\sim$133 K in HgBa$_2$Ca$_2$Cu$_3$O$_{8+\delta}$ (Hg1223)\cite{Schilling}, which reaches $T_{\rm c}$=164 K under pressure~\cite{Gao,Chu,Ihara}. 
The discovery of a remarkably high $T_{\rm c}$ in the cuprate family has not only opened many possibilities for potential technical applications, but has also provided a challenging research subject for condensed-matter physics and material sciences. 

Despite more than a quarter of a century of research, there is still no universally accepted theory for the mechanism of high-$T_{\rm c}$ superconductivity (HTSC) in cuprates. 
The main controversy arises with regard to the origin of the attractive force for the formation of Cooper pairs, which leads to such a remarkably high SC transition.
In conventional superconductors with a relatively low $T_{\rm c}$, the phonon-mediated electron-electron interaction is the attractive force for the formation of Cooper pairs in the Bardeen-Cooper-Schrieffer (BCS) theory, which was established half a century ago~\cite{BCS}. 
The HTSC in cuprates emerges on a layered structure of a two dimensional CuO$_2$ lattice  when an antiferromagnetic Mott insulator is doped with hole carriers. 
The strong hybridization of Cu-$3d_{x^2-y^2}$ and O-$2p\sigma$ orbitals brings a large superexchange interaction $J_{\rm in}\sim 0.12$ eV ($\sim$1300 K) in the CuO$_2$ plane for nondoped cuprates~\cite{JinLSCO,JinYBCO1,JinYBCO2,TokuraJ}. 
Therefore, an intimate relationship between antiferromagnetism (AFM) and HTSC is believed to be a key for understanding the origin of the remarkably high SC transition in cuprate superconductors\cite{Anderson,Anderson1,Anderson2,Chen,Giamarchi,Inaba,Lee1,Zhang,Himeda,Kotliar,Paramekanti1,Lee2,Demler,Yamase1,Yamase2,Paramekanti2,Shih1,Shih2,Senechal,Capone,Ogata,Pathak,Moriya}. 
Experimentally, however, in prototype high-$T_{\rm c}$ cuprates La$_{2-x}$Sr$_x$CuO$_4$ (LSCO),  AFM and HTSC are separated by the spin-glass phase~\cite{Keimer}. 
In LSCO, the chemical substitution of Sr$^{2+}$ for La$^{3+}$ is necessary to increase the planar CuO$_2$ hole density, but it introduces some local disorder simultaneously into nearly two-dimensional CuO$_2$ planes and buckling on a CuO$_{6}$ octahedral unit, which make doped hole carriers localize owing to the Anderson localization mechanism. 
As a result, it is inevitable that the intrinsic electronic characteristics of LSCO are masked so that AFM and HTSC are separated by the spin glass phase. 

\begin{figure*}[t]
\begin{center}
\includegraphics[width=15cm]{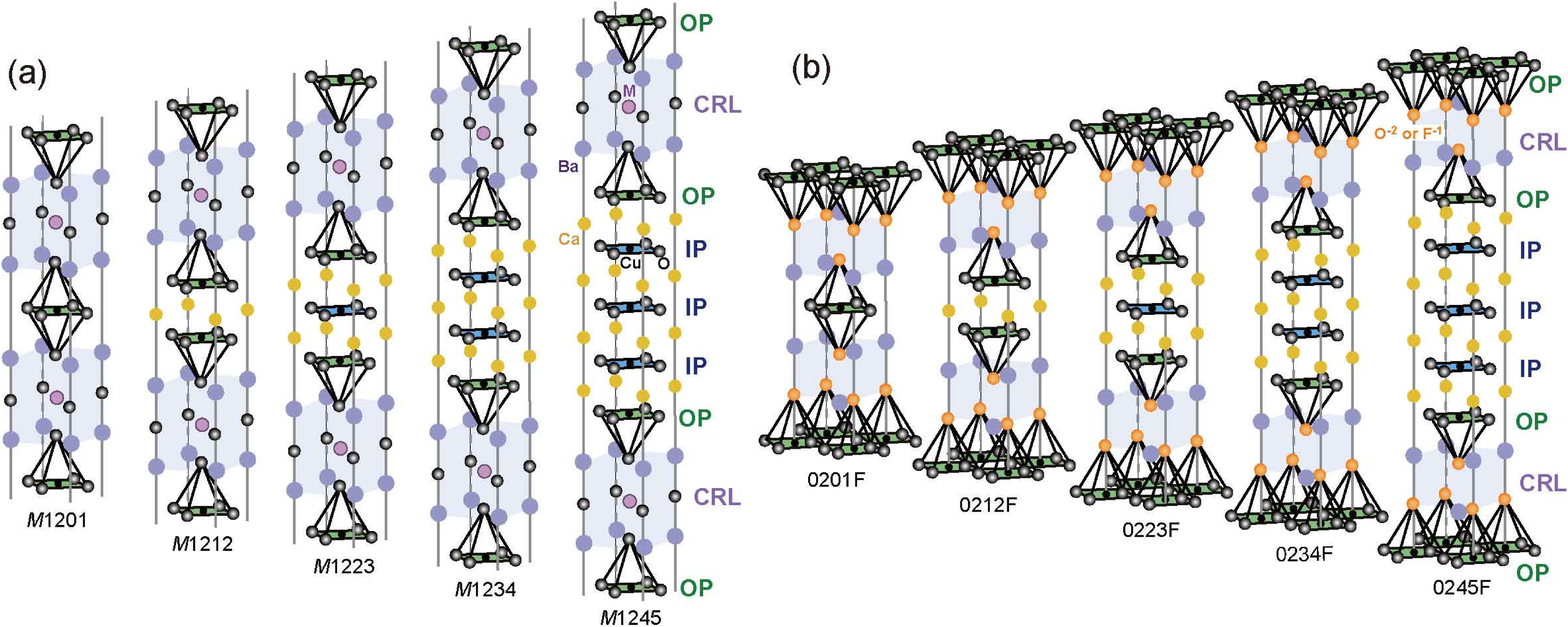}
\end{center}
\caption{(Color online) Crystal structure of $n$-layered cuprates of (a) $M$12($n$-1)$n$ and (b) 02($n$-1)$n$F. Copper oxides with more than three layers comprise inequivalent types of CuO$_2$ layers: an outer plane (OP) in a five-fold pyramidal coordination and an inner plane (IP) in a four-fold square coordination. 
Although the disorder may be introduced along with the chemical substitution in charge reservoir layers (CRLs), it is effectively shielded out of OPs, and hence homogeneously hole-doped CuO$_2$ planes with ideal flatness are realized, especially at IP, which is ensured by the narrow NMR linewidths (see Fig.~\ref{fig:CuNMR_comparison}).}
\label{fig:structure}
\end{figure*}

Multilayered cuprates provide us with the opportunity to research on the characteristics of a disorder-free CuO$_2$ plane with  hole carriers  homogeneously doped. 
Figures \ref{fig:structure}(a) and  \ref{fig:structure}(b) respectively show the crystal structures of $n$-layered cuprates in the series of $M$Ba$_2$Ca$_{n-1}$Cu$_n$O$_{2n+2+\delta}$ ($M$=Hg, Tl, and Cu) and Ba$_2$Ca$_{n-1}$Cu$_n$O$_{2n}$(F$_y$O$_{1-y}$)$_2$, denoted as $M$12($n$-1)$n$ and 02($n$-1)$n$F. 
Here, $n$ is the number of CuO$_2$ planes within a unit cell. 
Copper oxides with more than three layers comprise inequivalent types of CuO$_2$ layers: an outer CuO$_2$ plane (OP) in a five-fold pyramidal coordination and an inner  CuO$_2$ plane (IP) in a four-fold square coordination.
Site-selective nuclear magnetic resonance (NMR) and nuclear quadrupole resonance (NQR) studies are unique tools for differentiating layer-dependent electronic characteristics microscopically~\cite{TokunagaJLTP,Tokunaga,Kotegawa2001,Kotegawa2004,MukudaPRL2006,MukudaJPSJ2006,Shimizu2007,Mukuda2007PhysC,Mukuda2008,Shimizu2009PRB,Shimizu2009JPSJ,Mukuda2010,Itohara,Shimizu2011JPSJ,Shimizu2011PRB,Shimizu2011_n3,Tabata,KitaokaJPCS2011,KitaokaIOP}. 
One of the remarkable features and advantages of multilayered cuprates is that the CuO$_2$ layers are very flat and homogeneously doped, which have been ensured by the narrowest NMR linewidth to date among the very high quality cuprates investigated thus far~(for example, see Fig.~\ref{fig:CuNMR_comparison}). 
In multilayered cuprates, the carrier densities are inequivalent between OP and IP owing to an imbalance in the Madelung potential at each CuO$_2$ plane.  
Namely, since IP is farther from the charge reservoir layer (CRL) than OP, the carrier density at IPs is always lower than that at OPs. 
Carrier density can be tuned by the oxygen deficiency in CRLs ($M$O$_\delta$) for $M$12($n$-1)$n$ or by the chemical substitution of F at apical oxygen sites for 02($n$-1)$n$F.  
Such chemical substitutions introduce some disorder in CRLs, which may be slightly mapped onto OP, but the disorder potential at IP can be effectively shielded owing to the presence of conducting OP, as deduced from the narrower NMR linewidth at IP than at OPs (for example, see Figs.~\ref{fig:Cuspectra} and \ref{fig:CuNMR_comparison}). 
In this context, ideally flat CuO$_2$ planes are realized especially at underdoped IP, differentiating multilayered cuprates from monolayered cuprate LSCO. 

In this paper, we review a decade of extensive NMR investigations of $n$-layered cuprates with $n$=3, 4, and 5, which have revealed the intimate relationship between AFM and HTSC for a disorder-free CuO$_2$ plane with hole carriers homogeneously doped~\cite{Kotegawa2004,MukudaPRL2006,MukudaJPSJ2006,Shimizu2007,Mukuda2007PhysC,Mukuda2008,Shimizu2009PRB,Shimizu2009JPSJ,Mukuda2010,Itohara,Shimizu2011JPSJ,Shimizu2011PRB,Shimizu2011_n3,Tabata,KitaokaJPCS2011,KitaokaIOP,Shimizu_n5}. 
The intrinsic phase diagram possesses the following features: The AFM metallic state is robust and coexists uniformly with the HTSC at a single CuO$_2$ plane in a region extending up to the optimally doped one. 
The critical carrier density $p_{c}$ at which the AFM order collapses decreases from 0.10 to 0.08 to 0.075 as the interlayer magnetic coupling becomes weaker when decreasing from $n$=5 to 4 to 3, respectively.  
This provides a reasonable explanation why the AFM order in $n$=1:LSCO and $n$=2:YBa$_2$Cu$_3$O$_{6+x}$(YBCO$_{6+x}$) collapses at carrier densities with $p_{c}$=0.02 and 0.055, respectively.
We reveal that the SC gap and $T_{\rm c}$ exhibit a maximum irrespective of $n$ at $p\sim$~0.16 just outside $p_c(M_{\rm AFM}$=$0)\sim0.14$, where the AFM moment ($M_{\rm AFM}$) inherent in the CuO$_2$ plane totally disappears in the ground state.
We highlight that the ground-state phase diagram of AFM and HTSC (see Fig.~\ref{Mvsp}) is in good agreement with the ground-state phase diagrams in terms of either the $t$-$J$ model~\cite{Chen,Giamarchi,Inaba,Anderson,Anderson1,Anderson2,Lee1,Himeda,Kotliar,Paramekanti1,Lee2,Yamase1,Yamase2,Paramekanti2,Shih1,Shih2,Ogata,Pathak}, or the Hubbard model in the strong-correlation regime~\cite{Senechal,Capone}. 
The results presented here demonstrate that the in-plane superexchange interaction $J_{\rm in}$ plays a vital role as a glue for Cooper pairs or mobile spin-singlet pairs, which will lead us to a coherent understanding why $T_{\rm c}$ is so high for hole-doped cuprates.

\section{Experimental}  

\subsection{Sample preparation and characterization}

Polycrystalline powder samples of $n$-layered cuprates, i.e., $M$Ba$_2$Ca$_{n-1}$Cu$_n$O$_{2n+2+\delta}$ ($M$=Hg, Tl, and Cu) and apical F-substituted Ba$_2$Ca$_{n-1}$Cu$_n$O$_{2n}$(F$_y$O$_{1-y}$)$_2$, were prepared by the high-pressure synthesis technique, as described in the literature\cite{Tokiwa1,Tokiwa2,Iyo_TcVsn,Iyo1,Iyo2,Iyo_Hg_F,Shirage,Hirai}. 
To obtain more underdoped samples of the Hg1245 system, as-prepared samples in nearly optimally doped region [Hg1245(OPT)] were annealed in a quartz tube with Cu powder or in Ar gas atmosphere for more than several hundreds of hours\cite{Hirai,MukudaPRL2006,Mukuda2010}. 
In Ba$_2$Ca$_{n-1}$Cu$_n$O$_{2n}$(F$_y$O$_{1-y}$)$_2$, the substitution of oxygen (O$^{-2}$) for apical fluorine (F$^{-1}$), i.e., a decrease in nominal fluorine content ($y$), results in the doping of holes into CuO$_2$ layers, increasing $T_{\rm c}$\cite{Iyo_TcVsn,Iyo1,Iyo2,Iyo_Hg_F,Shirage}. 
Although the real apical fluorine F$^{-1}$ content may deviate slightly from the nominal one,  $T_{\rm c}$ and carrier density ($p$) can be tuned by changing the nominal content $y$ in this series, which provides an opportunity to investigate the characteristics of CuO$_2$ layers over a wide $p$ range systematically in the homologous series of $n$-layered cuprates \cite{Shimizu2009JPSJ,Shimizu2011_n3}.

\begin{figure}[tbp]
\begin{center}
\includegraphics[width=7cm]{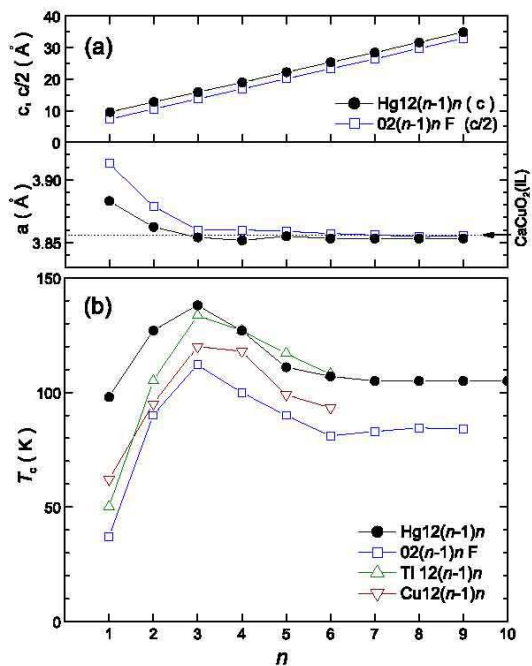}
\end{center}
\caption{(Color online) (a) Lattice parameters $a$ and $c$ plotted against  $n$  for the Hg12($n$-1)$n$ and 02($n$-1)$n$F systems\cite{Iyo_TcVsn,Iyo_Hg_F}. Note that the $c/2$ of 02($n$-1)$n$F is compared with the $c$ of Hg12($n$-1)$n$, because of the difference in the unit cell. 
(b) Relationship between $T_{\rm c}$ and $n$ for the homologous series of $n$-layered cuprates $M$12($n$-1)$n$ and 02($n$-1)$n$F.
 [cited from refs.\cite{Iyo_TcVsn,Iyo_Hg_F}] }
\label{fig:Tc_n}
\end{figure}

Powder X-ray diffraction measurements indicate that the samples used for NMR/NQR measurements are almost entirely composed of a single phase.  As shown in Fig. \ref{fig:Tc_n}(a), the $c$-axis length  monotonically increases with increasing $n$, which can be fitted using  linear functions, $c(n) \simeq 9.451+3.171\times(n-1)$[\AA] for Hg12($n$-1)$n$ and $c(n)/2 \simeq 7.205+3.191\times(n-1)$[\AA] for 02($n$-1)$n$F \cite{Iyo_TcVsn,Iyo_Hg_F}. 
The first term corresponds to the distance between OPs through CRL averaged in $n$-layered cuprates; the distances are  9.451[\AA] for Hg12($n$-1)$n$ and 7.205[\AA] for 02($n$-1)$n$F, originating from the difference in the structure of CRL between the two systems. 
The coefficient of the second term corresponds to the average distance between adjacent CuO$_2$ planes, which is almost equal to the $c$-axis length of the infinite layer CaCuO$_2$(IL) ($c(\infty)$= 3.179 \AA).
The $a$-axis length of two systems also approaches that of the CaCuO$_2$(IL) ($a(\infty)$= 3.856 \AA) as $n$ increases, as shown in the lower panel of Fig. \ref{fig:Tc_n}(a). 

The values of $T_{\rm c}$ of all the samples were uniquely determined by susceptibility measurement using a SQUID magnetometer, which exhibits a marked decrease due to the onset of SC diamagnetism. 
The $T_{\rm c}$ of nearly optimally doped samples exhibits a striking dependence on $n$, as shown in Fig. \ref{fig:Tc_n}(b), which has been confirmed in a homologous series\cite{Scott,Iyo_TcVsn,Iyo_Hg_F,Antipov}. 

\begin{figure}[tbp]
\begin{center}
\includegraphics[width=7cm]{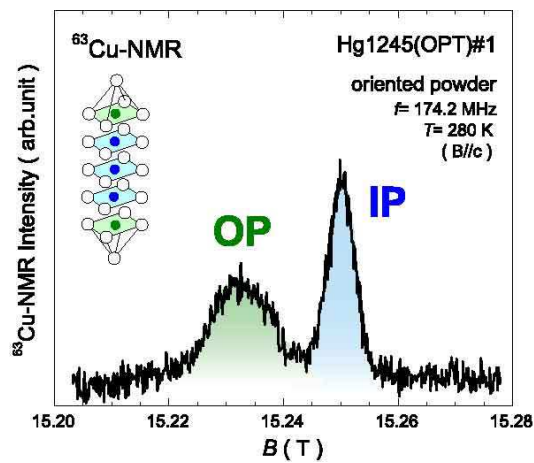}
\end{center}
\caption{(Color online) $^{63}$Cu-NMR spectrum of $n$=5:Hg1245(OPT)$\sharp1$. The $^{63}$Cu-NMR spectra at a pyramid-type outer CuO$_2$ plane (OP) and a square-type inner one (IP) are separately observed owing to the differences in Knight shift~\cite{Kotegawa2004}. The linewidths in the $^{63}$Cu-NMR spectra are as narrow as 50 Oe at IP and 110 Oe at OP at $\sim$15 T($B \| c$), indicating that IPs are ideally flat and homogeneously hole-doped.  [cited from ref.\cite{Kotegawa2004}] 
}
\label{fig:Cuspectra}
\end{figure}

\begin{figure}[tbp]
\begin{center}
\includegraphics[width=7cm]{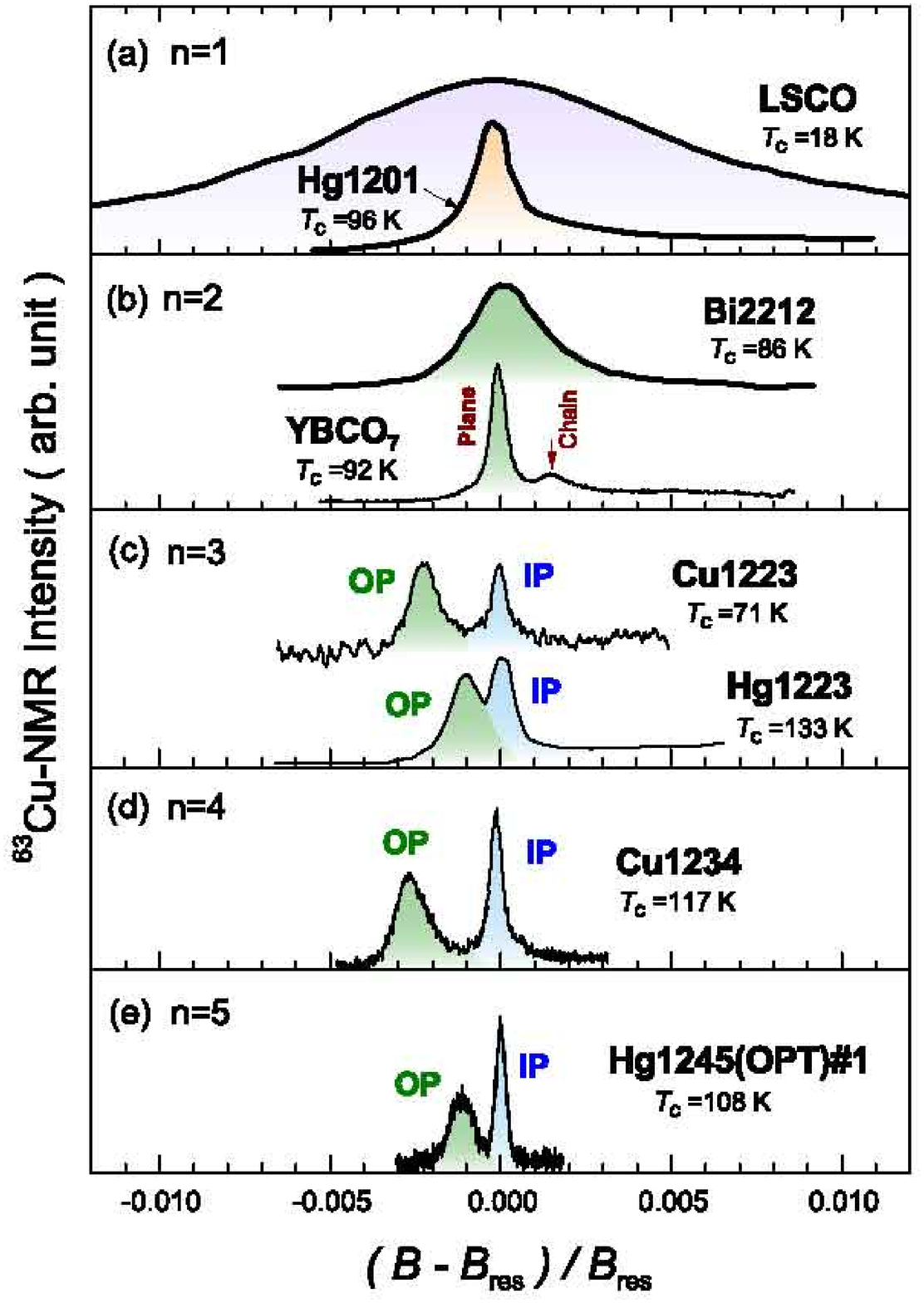}
\end{center}
\caption{(Color online) $^{63}$Cu-NMR spectra of (a) $n$=1: LSCO ($x$=0.24, $T_{\rm c}$=18 K)\cite{Ohsugi} and Hg1201 ($T_{\rm c}$=96 K)\cite{ItohHg1201_98}, (b) $n$=2: YBCO$_7$ ($T_{\rm c}$=92 K) and Bi2212 ($T_{\rm c}$=86 K)\cite{IshidaBi2212SG},  (c) $n$=3: Hg1223 ($T_{\rm c}$=133 K)\cite{MagishiPRB} and Cu1223 ($T_{\rm c}$=71 K)\cite{Kotegawa2001}, (d) $n$=4: Cu1234 ($T_{\rm c}$=117 K)\cite{Tokunaga,Kotegawa2001}, and (e) $n$=5: Hg1245(OPT)$\sharp1$ ($T_{\rm c}$=108 K)\cite{Kotegawa2004}. 
Here, the spectra are displayed against $(B-B_{\rm res})/B_{\rm res}$ for the normalization of differences in field conditions, where $B_{\rm res}$ is a resonance field for $^{63}$Cu. These spectra were measured under the following experimental conditions; LSCO ($B_{\rm res}\perp c$=10.79 T, $T$=20 K, aligned  polycrystal(APC))\cite{Ohsugi}, Hg1201 ($B_{\rm res}\parallel c$=7.46 T, $T$=100 K, APC)\cite{ItohHg1201_98}, YBCO$_7$ ($B_{\rm res}\parallel c$=15.08 T, $T$=90 K, single crystal), Bi2212 ($B_{\rm res}\parallel c$=10.95 T, $T$=77 K, single crystal)\cite{IshidaBi2212SG}, Hg1223 ($B_{\rm res}\parallel c$=10.95 T, $T$=140 K, APC)\cite{MagishiPRB}, Cu1223 ($B_{\rm res}\parallel c$=15.24 T, $T$=160 K, APC)\cite{Kotegawa2001}, Cu1234 ($B_{\rm res}\parallel c$=15.25 T, $T$=90 K, APC)\cite{Tokunaga,Kotegawa2001}, and Hg1245 ($B_{\rm res}\parallel c$=15.25 T, $T$=280 K, APC)\cite{Kotegawa2004}. 
}
\label{fig:CuNMR_comparison}
\end{figure}

\subsection{NMR/NQR measurements}

For NMR/NQR measurements, the powder samples were aligned along the $c$-axis at an external field ($B$) of $\sim$16 T and fixed using stycast 1266 epoxy. 
Figure \ref{fig:Cuspectra} shows a typical $^{63}$Cu-NMR spectrum of $n$=5:Hg1245(OPT)$\sharp1$, in which the spectra from OP and IP are separately observed owing to the difference in Knight shift, which enables us to study multilayered compounds site-selectively \cite{Julien,Magishi,MagishiPRB,
TokunagaJLTP,Tokunaga,Kotegawa2001,Kotegawa2004,MukudaPRL2006,MukudaJPSJ2006,Shimizu2007,Mukuda2007PhysC,Mukuda2008,Shimizu2009PRB,Shimizu2009JPSJ,Mukuda2010,Itohara,Shimizu2011JPSJ,Shimizu2011PRB,Shimizu2011_n3,Tabata,KitaokaJPCS2011,KitaokaIOP}. 
Owing to the difference in local structure between IP and OP, the $^{63}$Cu-NQR frequencies ($^{63}\nu_{Q}$) at OP and IP are typically $\sim$16 and $\sim$8.4 MHz, respectively, for $n$=5:Hg1245(OPT)$\sharp$1\cite{Kotegawa2004}, in which the presence of apical oxygen on OP generally makes the $^{63}\nu_{Q}$ of OP large. 
Moreover, it is remarkable that the linewidths in the $^{63}$Cu-NMR spectra are particularly narrow in multilayered cuprates, which are as narrow as 50 Oe for IP and 110 Oe for OP in $n$=5:Hg1245(OPT)$\sharp$1 even at $\sim$15 T($B \| c$). 
For comparison, the $^{63}$Cu-NMR spectra of typical $n$-layered cuprates are presented in Fig. \ref{fig:CuNMR_comparison}: (a) $n$=1 : LSCO ($x$=0.24, $T_{\rm c}$=18 K)\cite{Ohsugi} and Hg1201 ($T_{\rm c}$=96 K)\cite{ItohHg1201_98}, (b) $n$=2: YBCO$_7$ ($T_{\rm c}$=92 K) and Bi2212 ($T_{\rm c}$=86 K)\cite{IshidaBi2212SG}, (c) $n$=3: Hg1223 ($T_{\rm c}$=133 K)\cite{MagishiPRB} and Cu1223 ($T_{\rm c}$=71 K)\cite{Kotegawa2001}, (d) $n$=4: Cu1234 ($T_{\rm c}$=117 K)\cite{Tokunaga,Kotegawa2001}, and (e) $n$=5: Hg1245(OPT)$\sharp1$ ($T_{\rm c}$=108 K)\cite{Kotegawa2004}. 
Here, the spectra are displayed against $(B-B_{\rm res})/B_{\rm res}$  for the normalization of differences in field conditions, where the $B_{\rm res}$ is the resonance field for $^{63}$Cu. 
The broadening of $^{63}$Cu-NMR linewidth originates from the inhomogeneity of the Knight shift, and the distribution of the quadrupole shift when  CuO$_2$ planes are buckled. 
Thus, the figure indicates that the $^{63}$Cu-NMR linewidth becomes narrower as $n$ increases, suggesting that multilayered cuprates possess very flat CuO$_2$ layers from a microscopic point of view as well as homogeneous electronic states over the CuO$_2$ plane with hole carriers. 
Moreover, we also note that the SC in Hg1201($n$=1) and YBCO$_7$($n$=2) compounds also occurs on the CuO$_2$ plane with less disorder, which is one of the key factors for the relatively high SC transition at $T_{\rm c}$=96 and 92 K among $n$=1 and 2 compounds, respectively\cite{Eisaki}. 
These facts that the CuO$_2$ planes are ideally flat and homogeneously hole-doped in multilayered cuprates enable us to investigate the intrinsic properties of an ideal CuO$_2$ plane. 

The Knight shift $K$ generally comprises the temperature ($T$)-dependent spin part  $K_{\rm s}$  and the $T$-independent orbital part $K_{\rm orb}$ as follows: 
\begin{equation}
K^\alpha = K_{\rm s}^\alpha(T) + K_{\rm orb}^\alpha ~~~ (\alpha = c,ab), 
\end{equation}
where $\alpha$ is the direction of $B$.
The spin part of the Knight shift for the $B\parallel$ $ab$-plane ($K_{\rm s}^{ab}$) is obtained by subtracting   $K_{\rm orb}^{ab}$, which is approximately 0.23($\pm$0.02)\% assuming $K_{\rm s}^{ab}\approx 0$ in the $T$=0 limit. 
For multilayered cuprates that exhibit an AFM order, $K^{ab}_s(T)$ at nonmagnetic OP shows an upturn below $T_{\rm N}$ due to the transferred  hyperfine magnetic field arising from AFM IP, as shown in the lower panel of Fig.~\ref{fig:summary}: $K_{\rm orb}^{ab}\simeq$0.23($\pm$0.02)\% was assumed to be the same as that of  multilayered cuprates in which all CuO$_2$ planes are in a paramagnetic state. 
Note that $K_{\rm orb}^{ab}$'s are not so different among high-$T_{\rm c}$ cuprates whether at IP or OP. \cite{Julien,Magishi,MagishiPRB,Barrett,ItohHg1201_98,IshidaBi2212SG}

\subsection{Hyperfine magnetic field in CuO$_2$ plane}

According to the Mila-Rice Hamiltonian \cite{MilaRice}, the spin Knight shift of Cu in the CuO$_2$ plane is expressed as 
\begin{equation}
K_{s}^{\alpha}(T) = (A_{\alpha}+4B)\chi_s(T) ~~~(\alpha =c,ab), 
\label{eq:Ks}
\end{equation}
where $A_{\alpha}$ and $B$ are the on-site and supertransferred hyperfine fields of Cu, respectively. Here, $A_{\alpha}$ consists of  contributions induced by on-site Cu $3d_{x^2-y^2}$ spins $-$ anisotropic dipole, spin-orbit, and isotropic core polarization, and the $B$ term originates from the isotropic $4s$ spin polarization produced by four neighboring Cu spins through the Cu($3d_{x^2-y^2}$)-O($2p\sigma$)-Cu($4s$) hybridization. 
Since the spin susceptibility $\chi_s(T)$ is assumed to be isotropic, the anisotropy $\Delta$ of $K_{s}^{\alpha}(T)$ is given by 
\begin{equation}
\Delta \equiv \frac{K_{s}^{c}(T)}{K_{s}^{ab}(T)}=\frac{A_{c}+4B}{A_{ab}+4B}. 
\label{eq:B}
\end{equation}
The on-site hyperfine fields $A_{ab}$ $\approx$ 3.7 T/$\mu_{\rm B}$ and $A_c$ $\approx$ $-$17 T/$\mu_{\rm B}$ \cite{Monien,Millis,Imai} are assumed as material-independent in hole-doped high-$T_{\rm c}$ cuprates.  
In multilayered compounds [Hg1245(OPT)$\sharp1$], $B({\rm IP})\approx$ 6.1 T/$\mu_{\rm B}$ and  $B({\rm OP})\approx$ 7.4 T/$\mu_{\rm B}$ are estimated~\cite{Kotegawa2004}, which are larger than $B\sim$ 4 T/$\mu_{\rm B}$ for LSCO \cite{Ohsugi}, YBCO$_7$~\cite{Walstedt,Barrett,Takigawa}, and YBa$_2$Cu$_4$O$_8$(Y1248)~\cite{Zimmermann} compounds.
In the AFM ordered state where spins align antiferromagnetically among nearest-neighbor Cu sites in a two-dimensional lattice of the CuO$_2$ plane, the internal field at the Cu site is generally given by $B_{\rm int}$=$|A_{ab}-4B|M_{\rm AFM}$ from the AFM moment ($M_{\rm AFM}$) through those hyperfine interactions.

\subsection{Evaluation of planar CuO$_2$ hole density}

In hole-doped high-$T_{\rm c}$ cuprates, the hole density $p$ in CuO$_2$ planes has been determined by various methods: indirect chemical methods like solid solutions \cite{Shafer,Torrance,Kishino,Tokura}, bond valence sums determined from structural bond lengths \cite{Brown,TallonBV,Cava}, or the Fermi surface topology \cite{Kordyuk}. Besides, the thermoelectric power is a universal function of $p$ \cite{Obertelli,Tallon}, and the phase diagram for hole-doped cuprates is well described by $T_{\rm c}(p)$=$T_{\rm c}^{\rm max}[1-82.6(p-0.16)^2]$ \cite{Presland}, which are applicable to the estimation of $p$ when no suitable structural data are available.  
These methods are, however, inapplicable to multilayered cuprates composed of more than two inequivalent CuO$_2$ planes in a unit cell, because these methods evaluate the {\it total hole density}  not {\it each hole density} inherent in CuO$_2$ planes. 

\begin{figure}[tbp]
\begin{center}
\includegraphics[width=7.5cm]{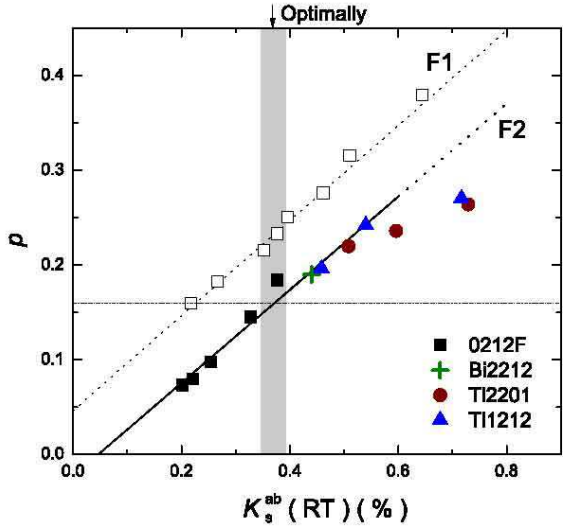}
\end{center}
\caption{\footnotesize (Color online) Plots of $K_{\rm s}^{ab}$(RT)s for 0212F\cite{Shimizu2011PRB}, Bi$_2$Sr$_2$CaCu$_2$O$_8$~(Bi2212),\cite{IshidaBi2212} Tl$_2$Ba$_2$CuO$_{6+\delta}$~(Tl2201),\cite{KitaokaTl2201} and TlSr$_2$CaCu$_2$O$_{7-\delta}$~(Tl1212),\cite{MagishiTl1212} as functions of $p$ evaluated using $T_{\rm c}$=$T_{\rm c}^{\rm max}[1-82.6(p-0.16)^2]$.\cite{Presland} 
Here, the above cuprates are selected because their local structures are homologous to those of the multilayered 02($n$-1)$n$F and $M$12($n$-1)$n$ series, which guarantees that the plot of $K_{\rm s}^{ab}$(RT) vs $p$ is material-independent among the homologous compounds. 
The relationship F2: ($p$=0.492$K_{\rm s}^{ab}$(RT)-0.023)~(solid line) between $K_{\rm s}^{ab}$(RT) and $p$ obtained by  fitting with the data for $K_{\rm s}^{ab}$(RT)$<$ 0.5\% allows us to separately estimate $p$ inherent in IP and OP in multilayered cuprates. 
More details of the validity of this relation are described in the literature.\cite{Shimizu2011PRB} 
Note that the relation F1:$p'$=0.502$K_{\rm s}^{ab}$(RT)+0.0462 used in the previous studies had overestimated the hole density by 0.06$\sim$0.07.\cite{TokunagaJLTP,Tokunaga,Kotegawa2001,Kotegawa2004,MukudaPRL2006,MukudaJPSJ2006,Shimizu2007,Mukuda2007PhysC,Mukuda2008,Shimizu2009PRB,Shimizu2009JPSJ,Mukuda2010,Itohara} [cited from ref. \cite{Shimizu2011PRB}]
}
\label{fig:Ks_p}
\end{figure}

In NMR experiments, the spin part of the Knight shift at room temperature ($K_{\rm s}^{ab}$(RT))  increases with $p$ from the underdoped region to the overdoped region in hole-doped cuprates \cite{Kotegawa2001,Walstedt1990,Ohsugi,IshidaBi2212,FujiwaraJPSJ,MagishiTl1212,Storey}, suggesting that $K_{\rm s}^{ab}$(RT) is available to determine planar CuO$_2$ hole densities ($p$). 
The linear equation $p'$=0.502$K_{\rm s}^{ab}$(RT)+0.0462 has been reported, as indicated by the dotted line (F1) in Fig. \ref{fig:Ks_p} \cite{TokunagaJLTP,Kotegawa2001}, where $p'$ is derived from the NQR frequencies of Cu and O in CuO$_2$ planes~\cite{Zheng}. 
However, when noting that $K_{\rm s}$(RT) for optimally doped cuprates is empirically 0.35 - 0.39\%, an optimal doping level was evaluated to be $p'$=0.22$\sim$0.24\cite{MukudaPRL2006,Mukuda2008,Shimizu2009JPSJ}, which is relatively larger than the widely accepted optimal doping level in hole-doped cuprates, i.e., $p$ $\sim$ 0.16, as shown in the figure. 
This inconsistency is, in part, due to the calculation that connects NQR frequency to hole density\cite{Haase}. 

Recently, we have investigated the bilayered ($n$=2) apical-fluorine compounds Ba$_2$CaCu$_2$O$_4$(F,O)$_2$ ($n$=2:0212F) over a wide carrier density range\cite{Shimizu2011PRB}, which provides an opportunity to reexamine $p$ by comparing it with those well established in other bilayered compounds, exhibiting the maximum $T_{\rm c}$ at $p(T_{\rm c}^{\rm max})$$\sim$~0.16. Figure \ref{fig:Ks_p} shows plots of $K_{\rm s}^{ab}$(RT) for $n$=2:0212F, Bi$_2$Sr$_2$CaCu$_2$O$_8$ ($n$=2:Bi2212) \cite{IshidaBi2212}, Tl$_2$Ba$_2$CuO$_{6+\delta}$ ($n$=1:Tl2201) \cite{KitaokaTl2201}, and TlSr$_2$CaCu$_2$O$_{7-\delta}$ ($n$=2:Tl1212) \cite{MagishiTl1212} as a function of $p$ evaluated using $T_{\rm c}(p)$=$T_{\rm c}^{\rm max}[1-82.6(p-0.16)^2]$ \cite{Presland}, indicating that $K_{\rm s}^{ab}$(RT) monotonically increases with $p$ from the underdoped region to the overdoped region. 
Here, the above cuprates are selected because their local structures are homologous to those of the multilayered 02($n$-1)$n$F and $M$12($n$-1)$n$ series, which guarantees that the plot of $K_{\rm s}^{ab}$(RT) vs $p$ is material-independent among these homologous compounds. 
More details of the validity of this relation have been described in the literature\cite{Shimizu2011PRB}. 
This renewed relation, based on the Knight shift, enables us to separately estimate $p$ for each CuO$_2$ plane in multilayered cuprates. 
In this review, we use the linear function of $p$=0.492$K_{\rm s}^{ab}$(RT)-0.023, as shown by the solid line (F2) in Fig. \ref{fig:Ks_p}, which was obtained by fitting with the data for $K_{\rm s}^{ab}$(RT)$<$ 0.5\%, since the $K_{\rm s}^{ab}$(RT) of the samples used in this study is less than 0.5\%. 
Note that the relation (F1) used in the previous studies had  overestimated the hole density by 0.06$\sim$0.07.\cite{TokunagaJLTP,Tokunaga,Kotegawa2001,Kotegawa2004,MukudaPRL2006,MukudaJPSJ2006,Shimizu2007,Mukuda2007PhysC,Mukuda2008,Shimizu2009PRB,Shimizu2009JPSJ,Mukuda2010,Itohara} 

According to eq.~(\ref{eq:Ks}), the $p$ dependence of $K_{s}^{ab}$(RT) is derived from those of the $B$ term in the hyperfine coupling constant and $\chi_s$(RT). In multilayered cuprates, the $B$ term increases moderately with $p$, showing a steep increase at $p$=0.18$\sim$0.20\cite{Shimizu2011PRB}. 
The $B$ term arises from Cu($3d_{x^2-y^2}$)-O($2p\sigma$)-Cu($4s$) covalent bonds with four nearest-neighbor Cu sites; therefore, the large $B$ suggests the strong hybridization between the Cu($3d_{x^2-y^2}$) and O($2p\sigma$) orbits. 
This result is consistent with the fact that a metallic state is more stabilized in an overdoped regime. 
For layers with tetragonal symmetry in cuprates homologous to the multilayered series, the $p$-dependent $B$ terms are {\it material-independent}\cite{Shimizu2011PRB}, being larger than the $B$ terms for LSCO, YBCO$_{6+x}$, and Y1248, $\sim$ 4 T/$\mu_{\rm B}$\cite{IshidaBi2212SG}.
This difference is also seen in the variation in the nuclear quadrupole frequency $^{63}\nu_Q$. 
$^{63}\nu_Q$ increases with $p$ for all the materials\cite{Ohsugi,Zheng,Haase}, while, for a given $p$, the absolute values of what LSCO and YBCO$_{6+x}$ are about 2 to 3 times larger than those for others\cite{Shimizu2011PRB}. 
$^{63}\nu_Q$ depends on the hole number $n_d$ in a Cu($3d_{x^2-y^2}$) orbit and on $n_p$ in a O($2p\sigma$) orbit. 
Therefore, it is expected that $n_d$ and $n_p$ in LSCO and YBCO$_{6+x}$ will be different from those in the other compounds treated here, even if they have the same $p$ ($p$=$n_d$+2$n_p-$1).
Actually, it has been reported that $n_d$ is large in LSCO and YBCO$_{6+x}$, which is the reason for the large $\nu_Q$ \cite{Zheng,Haase}.  
In this context, the reason why the $B$ terms in LSCO, YBCO$_{6+x}$, and Y1248 deviate from those in other materials is the different partitions of holes in the Cu($3d_{x^2-y^2}$) and O($2p\sigma$) orbits, which is probably related to the crystal structures; LSCO, YBCO$_{6+x}$, and Y1248 have orthorhombic crystal structures in the superconducting region, while 0212F, Bi2212, Tl2201, and Tl1212 have tetragonal ones. 
Thus, we conclude that the $p$ dependence of $B$ holds in CuO$_2$ planes with tetragonal symmetry homologous to the multilayered series, which guarantees the estimation of $p$ for OP and IP independently in multilayered cuprates using the renewed relation between $K_{s}^{ab}$(RT) and $p$.

\begin{figure*}[t]
\begin{center}
\includegraphics[width=15cm]{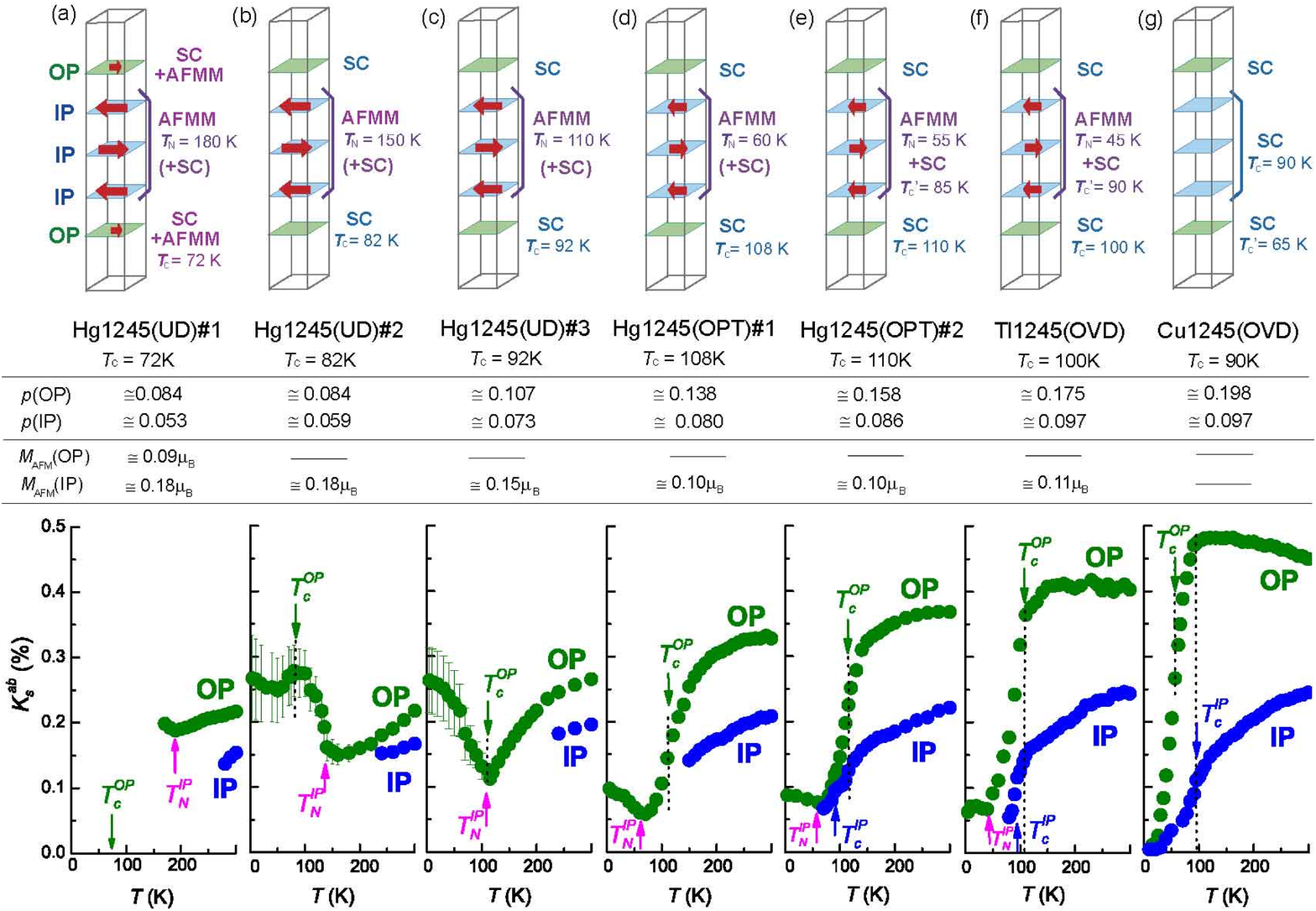}
\end{center}
\caption[]{\footnotesize (Color online) 
Illustration of layer-dependent physical properties for $n$=5 compounds:(a) Hg1245(UD)$\sharp1$ \cite{MukudaPRL2006,Tabata}, (b) Hg1245(UD)$\sharp2$\cite{Mukuda2010,Tabata},(c) Hg1245(UD)$\sharp3$ \cite{Mukuda2010,Tabata}, (d) Hg1245(OPT)$\sharp1$\cite{Kotegawa2004,Mukuda2008}, (e) Hg1245(OPT)$\sharp2$\cite{Mukuda2008}, (f) Tl1245(OVD)\cite{Kotegawa2004,Mukuda2008}, and (g) Cu1245(OVD)\cite{Kotegawa2001}. The middle panels present tables of the hole densities of $p$(IP) and $p$(OP), and the AFM ordered moments of $M_{\rm AFM}$(IP) and $M_{\rm AFM}$(OP).  The lower panels show the $T$ dependences of $K_{\rm s}^{ab}(T)$s, which enable us to separately estimate $p$s for IP and OP, and to probe the onset of AFM and HTSC at IP and OP.
}
\label{fig:summary}
\end{figure*}

\section{Results}

\subsection{Five-layered ($n$=5) compounds}

Figures \ref{fig:summary}(a) - \ref{fig:summary}(g) respectively show the layer-dependent physical properties unraveled by site-selective NMR studies of the following $n$=5 compounds: Hg1245(UD)$\sharp$1 ($T_{\rm c}$=72 K) \cite{MukudaPRL2006,Tabata}, Hg1245(UD)$\sharp$2 ($T_{\rm c}$=82 K) \cite{Mukuda2010,Tabata}, Hg1245(UD)$\sharp$3 ($T_{\rm c}$=92 K) \cite{Mukuda2010,Tabata}, Hg1245(OPT)$\sharp$1 ($T_{\rm c}$=108 K) \cite{Kotegawa2004,Mukuda2008}, Hg1245(OPT)$\sharp$2 ($T_{\rm c}$=110 K) \cite{Mukuda2008}, Tl1245(OVD) ($T_{\rm c}$=100 K) \cite{Kotegawa2004,Mukuda2008}, and Cu1245(OVD) ($T_{\rm c}$=90 K) \cite{Kotegawa2001}.
The temperature ($T$) dependences of $K^{ab}_s(T)$  are shown in the lower panel of the figures. 
The $K^{ab}_s$(RT)s at room temperature decrease with decreasing $p$, which enables us to evaluate $p$ at each CuO$_2$ plane using the relationship of $K^{ab}_s$(RT) vs $p$~\cite{Shimizu2011PRB}.

\subsubsection{Superconducting characteristics}

\begin{figure}[tbp]
\begin{center}
\includegraphics[width=7.5cm]{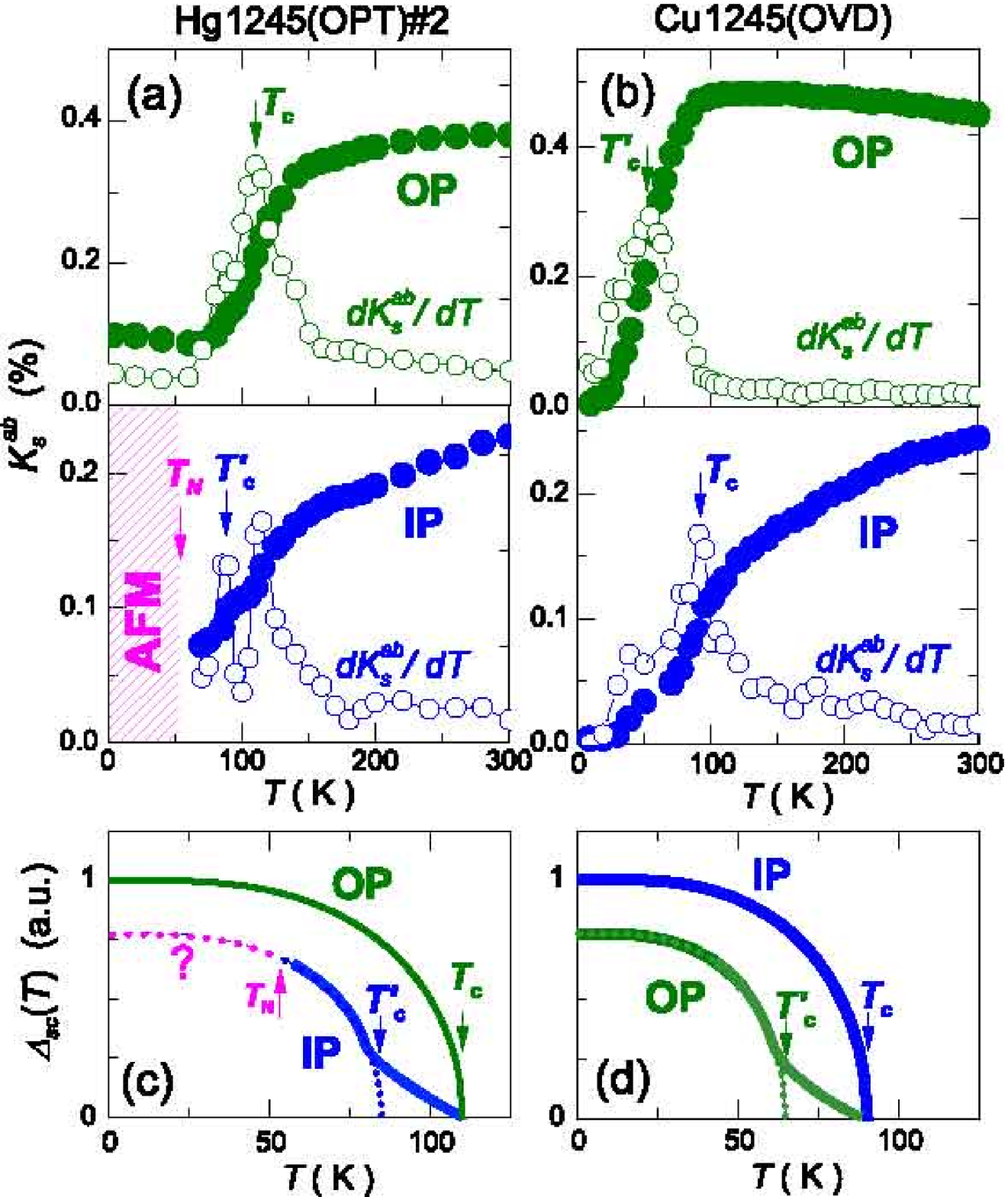}
\end{center}
\caption{(Color online) 
$T$ dependences of $K_{\rm s}^{ab}$s (solid circles) and its $T$ derivatives (empty circles) at OP and IP of (a) Hg1245(OPT)$\sharp2$ with $T_{\rm c}$=110 K and (b) Cu1245(OVD) with $T_{\rm c}$=90 K. Owing to a large imbalance in $p$s , $T_{\rm c}$s inherent in OP and IP are assigned from a peak in the $T$ derivatives of $K_{\rm s}^{ab}$ (see text). (a) A distinct peak in the $T$ derivatives of $K_{\rm s}^{ab}$ at OP coincides with the bulk $T_{\rm c}$= 110 K, demonstrating that the SC is driven by OPs. Another peak at $T$=85 K in the $T$ derivatives of $K_{\rm s}^{ab}$ at IP was assigned to $T_{\rm c}'$ inherent in IP.\cite{Mukuda2008} (b) A distinct peak in the $T$ derivatives of $K_{\rm s}^{ab}$ at IP coincides with the bulk $T_{\rm c}$= 90 K, demonstrating that the SC is driven by IPs.  A distinct peak at $T$=65 K in the $T$ derivatives of $K_{\rm s}^{ab}$ at OP was assigned to $T_{\rm c}'$ inherent in overdoped OP.\cite{TokunagaJLTP,Tokunaga,Kotegawa2001} (c,d) The SC gaps at IP (OP) for Hg1245(OPT)$\sharp2$~(Cu1245(OVD)) with $T_{\rm c}'$ lower than bulk $T_{\rm c}$ are anticipated to develop linearly between $T_{\rm c}$ and $T_{\rm c}'$ owing to the proximity effect.\cite{Tokunaga} [cited from refs.\cite{Kotegawa2001,Mukuda2008}] 
}
\label{fig:Knightshift}
\end{figure}

In multilayered cuprates, $T_{\rm c}$s inherent in OP and IP are estimated from the $T$ dependences of $K_{\rm s}^{ab}$ and its $T$ derivatives~\cite{TokunagaJLTP,Tokunaga,Kotegawa2001,Mukuda2008}, which are shown in Figs.~\ref{fig:Knightshift}(a) and \ref{fig:Knightshift}(b).
Figure~\ref{fig:Knightshift}(a) shows that a distinct peak in the $T$ derivatives of $K_{\rm s}^{ab}$ at OP coincides with the bulk $T_{\rm c}$= 110 K in optimally doped Hg1245(OPT)$\sharp2$, demonstrating that the SC is driven by OPs with an optimum $p$, while another peak at $T$=85 K in the $T$ derivatives of $K_{\rm s}^{ab}$ at IP is assigned to $T_{\rm c}'$ inherent to IP with a hole density smaller than that at OP~\cite{Mukuda2008}. 

Figure~\ref{fig:Knightshift}(b) shows that a distinct peak in the $T$ derivatives of $K_{\rm s}^{ab}$ at IP coincides with the bulk $T_{\rm c}=$ 90 K in overdoped Cu1245(OVD), demonstrating that the SC is driven by IPs, while a distinct peak at $T=65$ K in the $T$ derivatives of $K_{\rm s}^{ab}$ at OP was assigned to $T_{\rm c}'$ inherent in overdoped OP.~\cite{TokunagaJLTP,Tokunaga,Kotegawa2001} 
When noting that OP and IP are alternatively stacked along the $c$-axis in multilayered cuprates, the SC gaps at IP (OP) for Hg1245(OPT)$\sharp2$ [Cu1245(OVD)] with $T_{\rm c}'$ lower than bulk $T_{\rm c}$ are anticipated to develop linearly between $T_{\rm c}$ and $T_{\rm c}'$, as shown in Fig.~\ref{fig:Knightshift}(c) [Fig.~\ref{fig:Knightshift}(d)], respectively, owing to the proximity effect.  
Likewise, it was widely established that $T_{\rm c}$ in multilayered cuprates is determined by the carrier density at each CuO$_2$ plane~\cite{TokunagaJLTP,Tokunaga,Kotegawa2001,Mukuda2008,Shimizu2009JPSJ,Shimizu2011_n3}, and it was proposed that such two-gap SC causes anomalous behaviors~\cite{Hirai,YTanaka,Crisan,YTanaka_soliton}.

\subsubsection{Estimation of AFM moments $-$ zero-field Cu-NMR/NQR studies}

\begin{figure}[tbp]
\centering
\includegraphics[width=7.5cm]{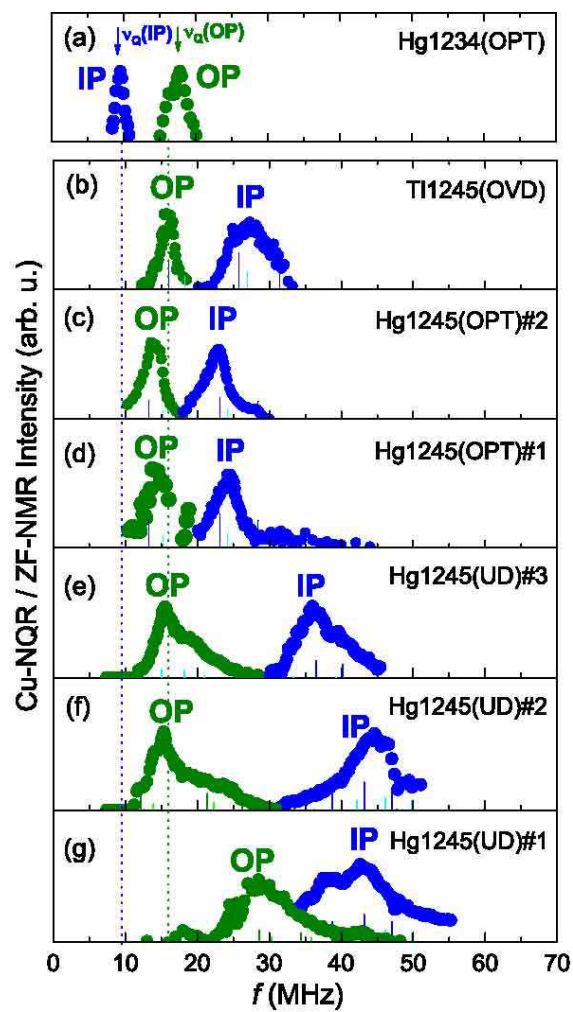}
\caption[]{\footnotesize (Color online) Cu-NQR/zero-field NMR spectra at 1.5 K for (b) Tl1245(OVD),\cite{Kotegawa2004,Mukuda2008} (c) Hg1245(OPT)$\sharp2$,\cite{Mukuda2008} (d) Hg1245(OPT)$\sharp1$,\cite{Kotegawa2004,Mukuda2008} (e) Hg1245(UD)$\sharp3$,\cite{Tabata} (f) Hg1245(UD)$\sharp2$,\cite{Tabata} and (g) Hg1245(UD)$\sharp1$\cite{MukudaPRL2006,Tabata,IP_UD}, along with (a) $n$=4:Hg1234(OPT) with $T_{\rm c}$=123 K  in the paramagnetic state.\cite{Itohara}  Dotted lines represent the NQR frequencies $^{63}\nu_{Q}$(IP)=8.4 and $^{63}\nu_{Q}$(OP)=16 MHz for Hg1245(OPT)$\sharp1$.\cite{Kotegawa2004} The solid bars represent resonance frequencies and intensities for two components of isotopes, $^{63}$Cu and $^{65}$Cu, which were calculated based on eq.~(\ref{eq:hamiltonian}). [cited from refs.\cite{Itohara,Kotegawa2004,MukudaPRL2006,Mukuda2008,Tabata}] 
}
\label{fig:ZFNMR}
\end{figure}

The observation of a zero-field (ZF) Cu-NMR spectrum at low $T$ allows us to estimate the AFM ordered moment $M_{\rm AFM}$. 
In general, the Hamiltonian for Cu nuclear spins ($I=3/2$) in crystal lattices with an axial symmetry is described by the Zeeman interaction due to the magnetic field ${\bm B}$, ${\cal H}_{Z}$, and the nuclear-quadrupole interaction ${\cal H}_{Q}$ as
\begin{eqnarray}
{\cal H}&=&{\cal H}_Z+{\cal H}_Q  \notag \\
        &=&-\gamma_N \hbar {\bm I} \cdot {\bm B}+\frac{e^{2}qQ}{4I(2I-1)}(3I_{z^{\prime}}^2-I(I+1)),
\label{eq:hamiltonian}
\end{eqnarray}
where $\gamma_{N}$ is the Cu nuclear gyromagnetic ratio, $eQ$ is the nuclear quadrupole moment, and $eq$ is the electric field gradient at the Cu nuclear site. In ${\cal H}_{Q}$, the nuclear quadrupole resonance (NQR) frequency is defined as $\nu_{Q}=e^{2}qQ/2h$. 
In nonmagnetic substances, the NQR spectrum originates from the second term in eq.~(\ref{eq:hamiltonian}) at a zero external field ($B_{\rm ext}$=0 T). On the other hand, in magnetically ordered substances, the ZF-NMR spectrum is observed owing to the internal magnetic field $B_{\rm int}$ at Cu sites in eq. (\ref{eq:hamiltonian}) despite $B_{\rm ext}$=0 T.
Figure~\ref{fig:ZFNMR}(a) shows a typical Cu-NQR spectrum of $n$=4:Hg1234(OPT) with $T_{\rm c}$=123 K at 1.5 K\cite{Itohara}. The respective $^{63}$Cu-NQR frequencies ($^{63}\nu_{Q}$) at IP and OP are 9.6 and 17.8 MHz, which are comparable to the typical $^{63}\nu_{Q}$ values at IP and OP in a paramagnetic regime\cite{Julien,MagishiPRB}. 

Figures \ref{fig:ZFNMR}(b)-\ref{fig:ZFNMR}(g) respectively show the Cu-NQR/ZF-NMR spectra at 1.5 K for Tl1245(OVD)\cite{Kotegawa2004,Mukuda2008} Hg1245(OPT)$\sharp2$\cite{Mukuda2008}, Hg1245(OPT)$\sharp1$\cite{Kotegawa2004,Mukuda2008}, Hg1245(UD)$\sharp3$\cite{Tabata}, Hg1245(UD)$\sharp2$\cite{Tabata}, and Hg1245(UD)$\sharp1$\cite{MukudaPRL2006,Tabata}. In these spectra, no NQR spectrum at IP is observed, pointing to an onset of AFM order at IP, whereas the NQR spectrum at OP is observed for (b) Tl1245(OVD)\cite{Kotegawa2004,Mukuda2008}, (c) Hg1245(OPT)$\sharp2$\cite{Mukuda2008}, (d) Hg1245(OPT)$\sharp1$\cite{Kotegawa2004,Mukuda2008}, (e) Hg1245(UD)$\sharp3$, and (f) Hg1245(UD)$\sharp2$. 

These spectra at IP, that are observed in the range of 20$\sim$50 MHz, are reproduced by assuming the internal field $B_{\rm int}$ at IP, which is generally given by $B_{\rm int}$=$|A_{\rm hf}|M_{\rm AFM}$=$|A_{ab}-4B|M_{\rm AFM}$. 
Here, $A_{ab}\approx$ 3.7 T/$\mu_{\rm B}$, $B({\rm IP})\approx$ 6.1 T/$\mu_{\rm B}$ and  $B({\rm OP})\approx$ 7.4 T/$\mu_{\rm B}$ are assumed in multilayered compounds\cite{Kotegawa2004}, $M_{\rm AFM}$ is a spontaneous AFM moment at Cu sites.  Using these values, $M_{\rm AFM}({\rm IP})$s at IP are estimated to be in the range of 0.1$\sim$0.18 $\mu_{\rm B}$ at $T$=1.5 K, which are listed up in the middle panel of Fig.~\ref{fig:summary}.  Note that the Cu-NMR spectra in undoped Mott insulators are observed in the higher-frequency ranges, such as 75$\sim$110 MHz for La$_2$CuO$_4$\cite{Tsuda} and YBCO$_6$\cite{YasuokaZF,Tsuda} or 125$\sim$150 MHz for CaCuO$_2$(IL) (see Fig.\ref{fig:n8}(f)) from the $M_{\rm AFM}$=0.5$\sim$0.7$\mu_{\rm B}$\cite{Vaknin1987,Vaknin1989}. 
Thus, the mobile holes existing at IP uniformly reduce to $M_{\rm AFM}$(IP)=0.1$\sim$0.18 $\mu_{\rm B}$, indicating that a static AFM {\it metallic} state is realized at IP uniformly. 
Here, note that the presence of the spin-glass phase is excluded because the possible distribution in $M_{\rm AFM}({\rm IP})$ is less than $\pm$ 0.02$\mu_{\rm B}$. 

In the most underdoped Hg1245(UD)$\sharp$1, the spectrum at OP is observed at approximately 30 MHz (see Fig.~\ref{fig:ZFNMR}(g)), indicating that $M_{\rm AFM}$(OP)$\sim$ 0.092$\mu_{\rm B}$ even at OP that is responsible for the onset of HTSC with $T_{\rm c}$= 72 K. 
Both the phase separation and the spin-glass phase are excluded since no NQR spectra are observed. 
These facts indicate that both AFM and HTSC  uniformly coexist at the microscopic level\cite{MukudaPRL2006}. 
Here, note that the broadening of these spectra at OP and IP points to a possible distribution of $M_{\rm AFM}$(OP) and $M_{\rm AFM}$(IP) with approximately $\pm$ 0.02 $\mu_{\rm B}$ over the samples.  This may be partly because some subtle inhomogeneity in carrier density takes place owing to deoxidization process. 
Note that $M_{\rm AFM}$ and $T_{\rm c}$ at OP in Hg1245(UD)$\sharp$1  are comparable to those at IP in Hg1245(OPT)$\sharp$2, since the $p$ values are almost the same for these layers. 
This means that $M_{\rm AFM}$ and $T_{\rm c}$ are primarily determined  by $p$ at each layer not by the electronic states of adjacent layers, whereas $T_{\rm N}$ depends on the number $n$ of CuO$_2$ layers or on interlayer magnetic coupling, as discussed in $\S$4.1 and  $\S$4.2.

\subsubsection{Determination of N\'eel temperature $T_{\rm N}$ for AFM order}

\begin{figure}[tbp]
\begin{center}
\includegraphics[width=7.5cm]{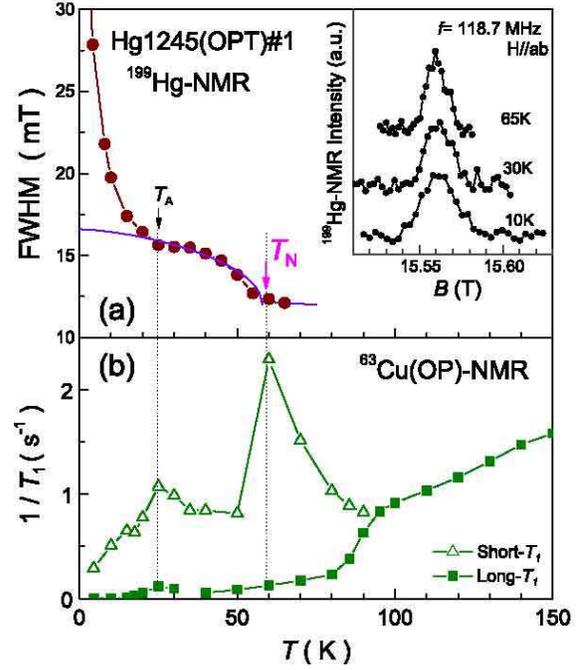}
\end{center}
\caption{(Color online) (a) $T$ dependences of $^{199}$Hg-NMR spectrum and its full-width at half-maximum (FWHM) when the $B \perp$ $c$-axis for $n$=5:Hg1245(OPT)$\sharp$1.  The solid line indicates the fitting of FWHM to the relation $M_{\rm AFM}(T)\propto (1-(T/T_N)^{3/2})^{1/2}$, which is the theoretical prediction for weak itinerant AFM metals\cite{MoriyaAFM}. 
(b) $T$ dependences of short (triangles) and long (squares) components in $^{63}$Cu-NMR $1/T_1$ at OP. A peak in $1/T_1$ points to $T_{\rm N}$=60 K, below which the FWHM at the Hg site  increases owing to the development of $M_{\rm AFM}$(IP).\cite{Kotegawa2004} 
[cited from ref.\cite{Kotegawa2004}]
}
\label{fig:spectraHg}
\end{figure}

The N\'eel temperature $T_{\rm N}$ is determined by the measurement of nuclear-spin-relaxation rate $1/T_{\rm 1}$, 
which exhibits a peak at $T_{\rm N}$. Generally, $1/T_{\rm 1}$ is described as
\begin{equation}
\frac{1}{T_{1}}=\frac{2\gamma_{\rm N}^{2}k_{\rm B}T}{(\gamma_{\rm e} \hbar )^{2}}\sum_{\bm q} |A_{\bm q}|^{2}\frac{{\rm Im}[\chi(\bm q, \omega_{0})]}{\omega_{0}},
\label{eq:T1}
\end{equation}
where $A_{\bm q}$ is a wave-vector (${\bm q}$)-dependent hyperfine-coupling constant, $\chi({\bm q},\omega)$ is the dynamical spin susceptibility, and $\omega_0$ is the NMR frequency. 
The $T$ dependence of $1/T_1$ exhibits a peak at $T_{\rm N}$ because low-energy spectral weight in $\chi({\bm q}={\bm Q},\omega)$ is strongly enhanced at $\omega_0 \sim 0$ in association with the divergence in magnetic correlation length at $T \sim T_{\rm N}$.  Here, ${\bm Q}$ is an AFM wave vector ($\pi$, $\pi$). In the case of Hg1245(OPT)$\sharp$1, for instance, the AFM order at IP was detected by $^{63}$Cu-NMR $1/T_1$ at OP that shows a peak at $T_{\rm N}$=60 K, as shown in Fig. \ref{fig:spectraHg}(b)\cite{Kotegawa2004}. Likewise, the $T_{\rm N}$s at IP were 55 and 45 K for Hg1245(OPT)$\sharp2$\cite{Mukuda2008} and Tl1245(OVD)\cite{Kotegawa2004,Mukuda2008,Mukuda2007PhysC}, respectively. Note that the AFM order below $T_{\rm N}=60$ K in  Hg1245(OPT)$\sharp$1 was also confirmed  by the $\mu$SR measurement\cite{muSR,muSR2}. Note that the onset of AFM order at IP was corroborated by an upturn in $K^{ab}_s(T)$ at OP upon cooling below the  $T_{\rm N}$s in Hg1245(OPT)$\sharp1$, Hg1245(OPT)$\sharp2$, and Tl1245(OVD), as marked by upward-arrows in the lower panel of Fig.~\ref{fig:summary}. 
As a result, the $T_{\rm N}$s for the more underdoped compounds such as Hg1245(UD)$\sharp$1, Hg1245(UD)$\sharp$2, and Hg1245(UD)$\sharp$3 were evaluated from the temperature below which the $K^{ab}_s(T)$ at OP reveals an upturn. As expected,  as $p$ decreases, the $T_{\rm N}$ at IP increases from $T_{\rm N}\sim$110, 150, and 180 K for Hg1245(UD)$\sharp$3, Hg1245(UD)$\sharp$2~\cite{Mukuda2010,Tabata}, and Hg1245(UD)$\sharp$1, respectively~\cite{Tabata}. 

The $T$ variation of $M_{\rm AFM}$ below $T_{\rm N}$ was indirectly probed on the basis of the $T$ dependence of the internal field at nuclear sites in charge reservoir layers. 
Figure \ref{fig:spectraHg}(a) shows the $T$ dependence of the $^{199}$Hg-NMR spectrum for the $B \perp c$-axis and its full-width at half-maximum (FWHM)  of Hg1245(OPT)$\sharp$1 with $T_{\rm c}$=108 K and $T_{\rm N}$=60 K. 
The FWHM increases rapidly below $T_{\rm N}$= 60 K, probing the development of the internal field at the $^{199}$Hg site induced by the onset of $M_{\rm AFM}$(IP). The $T$ dependence of $M_{\rm AFM}$(IP) was close to the theoretical prediction for  weak itinerant AFM metals \cite{MoriyaAFM}, as indicated by the solid line in Fig.~\ref{fig:spectraHg}(a). 
Note that the FWHM of the $^{199}$Hg-NMR spectrum is increased markedly below $T_{\rm A}$= 25 K, which has not been identified yet\cite{Kotegawa2004}.

\begin{figure}[tbp]
\begin{center}
\includegraphics[width=7.5cm]{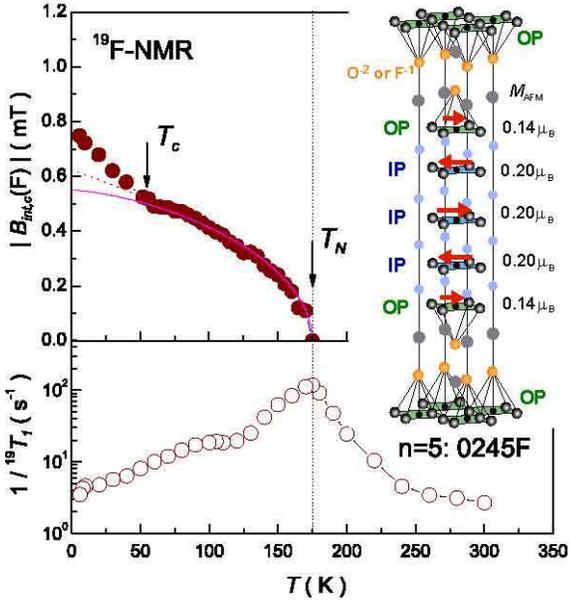}
\end{center}
\caption{(Color online) (a) $T$ dependence of the internal field at apical F site ($|B_{\rm int}^c({\rm F})|$) for $n$=5:0245F with $T_{\rm c}$=52 K and $T_{\rm N}$=175 K.\cite{Shimizu2011JPSJ} $|B_{\rm int}^c({\rm F})|$ was estimated from the splitting of the $^{19}$F-NMR spectra in the $B\parallel$ $c$-axis. (b) $T_{\rm N}$=175 K was determined by a peak of $^{19}$F-NMR $1/T_1$. The solid and dotted lines represent $B_{\rm int}^c$(F)$\propto (1-(T/T_N)^{3/2})^{(1/2)}$,\cite{MoriyaAFM} and $\propto(1-T/T_N)^{(1/2)}$,\cite{Borsa} respectively. [cited from ref.\cite{Shimizu2011JPSJ}]
}
\label{fig:Fdata}
\end{figure}

Recently, in the apical-F multilayered compound $n$=5:0245F with $T_{\rm c}$=52 K, which is more underdoped than $n$=5:Hg1245(UD)$\sharp$1 with $T_{\rm c}$=72 K, clear evidence of the $T$ evolution of $M_{\rm AFM}$ has been presented below $T_{\rm N}$=175 K, along with  an SC diamagnetic shift below $T_{\rm c}$=52 K through $^{19}$F-NMR studies\cite{Shimizu2011JPSJ}. 
The ZF Cu-NMR study revealed that $M_{\rm AFM}$(IP)=0.20 $\mu_{\rm B}$ and $M_{\rm AFM}$(OP)=0.14 $\mu_{\rm B}$ at 1.5 K. 
As shown in Fig.~\ref{fig:Fdata}, the internal field $|B_{\rm int}^c({\rm F})|$ at the apical-F site for the $B\parallel$ $c$-axis, which was evaluated from the splitting of the $^{19}$F-NMR spectra, increases significantly below $T_{\rm N}=$ 175 K \cite{HF_Fsite}, at which the $^{19}$F-NMR $1/T_1$ also exhibits a peak. 
The $T$ dependence of $|B_{\rm int}^c({\rm F})|$ at the apical-F site for $B\parallel$ $c$-axis was roughly reproduced down to $T_{\rm c}$= 52 K by either $B_{\rm int}^c$(F)$\propto$ $M_{\rm AFM}(T)$=$M_{\rm AFM}(0)(1-(T/T_N)^{3/2})^{1/2}$ (solid line)\cite{MoriyaAFM} or $\propto(1-(T/T_N))^{1/2}$ (dotted line) used for slightly doped LSCO compounds~\cite{Borsa}. 
These results convinced us of the three-dimensional long-range AFM order below $T_{\rm N}$=175 K in $n$=5:0245F with the SC transition below $T_{\rm c}$=52 K. 
Here, we note that $B_{\rm int}^c$(F) shows an additional increase as $T$ decreases below $T_{\rm c}$; the increase in $B_{\rm int}^c$(F) below $T_{\rm c}$ implies that the onset of an SC order parameter is actually coupled with $M_{\rm AFM}$(OP) in the HTSC-AFM coexisting state.\cite{Shimizu2011JPSJ} 
This result gives convincing evidence that the HTSC below $T_{\rm c}$=52 K emerges on the background of the AFM order taking place below $T_{\rm N}$=175 K; hence, both coexist.

\subsubsection{Pseudogap behavior in $n$=5 compounds}

Since the discovery of high-$T_{\rm c}$ cuprates, anomalous normal states exhibiting a pseudogap behavior have been one of the most important subjects in the research on HTSC. Initially, a gaplike behavior was reported as a gradual suppression of $1/T_1T$ below $T^{*}$~\cite{Yasuoka}, which has been called the {\it spin gap}. 
Neutron scattering experiments in underdoped regions also showed that spin excitations at low energies are suppressed in the normal state~\cite{Rossat}. Then, angle-resolved-photoemission spectroscopy (ARPES) experiments directly identified the existence of an energy gap in electronic spectra even above $T_{\rm c}$~\cite{Loeser,Ding}. 
This gap observed in ARPES below $T^{*}$ turned out to have the same angular dependence as the $d$-wave SC gap in the Brillouin zone~\cite{Ding,Harris}. 
After these experimental observations of single-particle spectra, the gap has been called the {\it pseudogap}. 

\begin{figure}[tbp]
\begin{center}
\includegraphics[width=7.5cm]{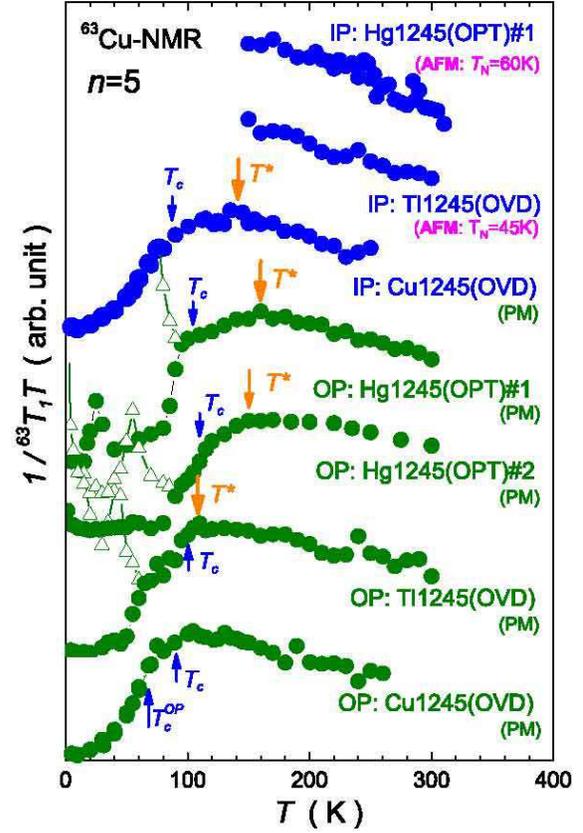}
\end{center}
\caption{(Color online) $T$ dependence of $^{63}$Cu-NMR $1/T_1T$ at IP and OP in $n$=5 compounds\cite{Kotegawa2001,Kotegawa2004,Mukuda2007PhysC,Mukuda2008}. The data are presented from the most underdoped IP (top) to the heavily overdoped OP (bottom).  The $T^{*}$ of $n$=5 compounds increases as $p$ decreases, but the underdoped IP, which shows the AFM order  below 60 K (Hg1245(OPT)$\sharp1$) and 45 K (Tl1245(OVD)), shows no indication of a pseudogap above 140 K. This suggests that the {\it spin-gap} collapses in the AFM-HTSC mixed state for $p < p_c(n)$.[cited from refs. \cite{Kotegawa2001,Kotegawa2004,Mukuda2007PhysC,Mukuda2008}] 
}
\label{fig:n5_PG}
\end{figure}

Figure \ref{fig:n5_PG} indicates the $T$ dependence of $^{63}$Cu-NMR $1/T_1T$ at IP and OP of the $n$=5 compounds for the $B\parallel$ $c$-axis\cite{Kotegawa2001,Kotegawa2004,Mukuda2007PhysC,Mukuda2008}.  The $1/T_1T$ at OP for Hg1245(OPT)$\sharp1$ starts to decrease upon cooling below $T^*$=160 K \cite{Kotegawa2004}. As shown in Fig.~\ref{fig:PhaseDiagram_n5}, $T^{*}$ decreases as $p$ increases as is also observed in single- and bilayered compounds~\cite{Yasuoka,REbook,IshidaBi2212,ZhengPG}. 
However, note that when the underdoped IPs of Hg1245(OPT)$\sharp1$ and Tl1245(OVD) show an AFM order below $T_{\rm N}$=60 K and 45 K, respectively, $1/T_1T$ exhibits no indication of a pseudogap behavior above 140 K. 
These results reveal that low-energy spectral weights in $\chi(Q,\omega)$ at IP of Hg1245(OPT)$\sharp1$ and Tl1245(OVD) are critically enhanced around $\omega\rightarrow0$ toward the AFM order. As a result, the NMR spectra at IP are lost below 140 K owing to the extremely short nuclear relaxation times.  By contrast, $1/T_1T$ at the slightly overdoped IP for Cu1245(OVD) shows the pseudogap state below $T^*$=145 K~\cite{Kotegawa2001,Kotegawa2004,Mukuda2007PhysC}. 
Eventually, we highlight that a {\it spin-gap} collapses in the underdoped region where the AFM order emerges.
Recently, a similar behavior on the $p$ dependence of $T^*$ has been observed in the $n$=3 compounds \cite{Shimizu2011_n3}.
This pseudogap behavior is also extensively discussed in $\S$ 4.4.

\subsubsection{Phase diagram in $n$=5 compounds}

\begin{figure}[tbp]
\begin{center}
\includegraphics[width=7.5cm]{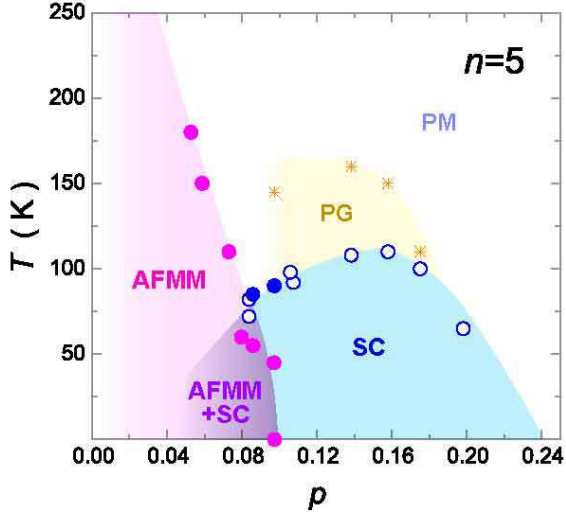}
\end{center}
\caption{(Color online) Phase diagram of AFM and HTSC at homogeneously doped CuO$_2$ plane in $n$=5 compounds.  $T_{\rm N}$, $T_{\rm c}$, and $T^*$ are plotted against the hole density $p$\cite{Shimizu2011PRB}. The solid and empty circles correspond to the data for IP and OP, respectively. 
The AFM metallic phase is robust and uniformly coexists with the HTSC state up to a quantum critical density $p_c(5)\sim$ 0.1 at which AFM order collapses. $T_{\rm c}$ exhibits a maximum at $p(T_{\rm c}^{\rm max})\sim$ 0.16. Here, PM and PG denote the paramagnetic phase and pseudogap state, respectively. The star denotes the pseudogap temperature $T^*$. [cited from refs.\cite{MukudaPRL2006,Mukuda2008,Shimizu2011PRB}]
}
\label{fig:PhaseDiagram_n5}
\end{figure}

Figure \ref{fig:PhaseDiagram_n5} shows a novel phase diagram for $n$=5 compounds. Here, $T_{\rm N}$ and $T_{\rm c}$ are plotted as functions of $p$, which is estimated from the relationship of $K_{\rm s}^{ab}$(RT) vs $p$ discussed in $\S$ 2.4.
The characteristic features are summarized as follows: (i) The AFM {\it metallic} phase (AFMM) is robust up to $p \sim$0.1 and uniformly coexists with the HTSC state up to a quantum critical density $p_c(n$=$5)\sim$ 0.1 at which the AFM order collapses. (ii) $T_{\rm c}$ has a peak at $p\sim$0.16 after the AFM order collapses. 
These findings suggest the intimate relationship between HTSC and AFM. 
(iii) The $p$ dependence of $T^*$ below which $1/T_1T$ starts to decrease points to that a {\it spin-gap} collapses in the AFM-HTSC mixed state. These features of a phase diagram differ significantly from the well-established phase diagram of LSCO \cite{Keimer}, in which the AFM and HTSC phases are separated by the spin-glass phase in association with the Anderson localization mechanism (see Fig. \ref{fig:PhaseDiagram}).

\subsection{Phase diagram in $n$=4 compounds}

\begin{figure}[tbp]
\begin{center}
\includegraphics[width=7.5cm]{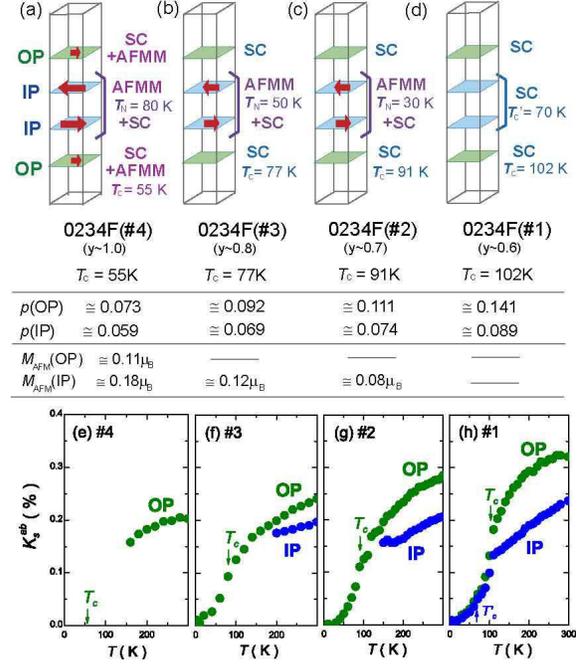}
\end{center}
\caption{(Color online) 
Illustration of layer-dependent physical properties for $n$=4:0234F: (a) $\sharp4(y\sim$1.0), (b) $\sharp3(y\sim$0.8), (c) $\sharp2(y\sim$0.7), and (d) $\sharp1(y\sim$0.6). The middle panels present tables of the hole densities of $p$(IP) and $p$(OP), and the AFM ordered moments of $M_{\rm AFM}$(IP) and $M_{\rm AFM}$(OP) (see text).  The lower panels show the $T$ dependences of $K_{\rm s}^{ab}(T)$s, which enable us to separately estimate $p$s for IP and OP, and to probe the onset of AFM and HTSC at IP and OP.
[cited from refs.\cite{Shimizu2009JPSJ,Shimizu2011PRB}]  
}
\label{fig:summary_n4}
\end{figure}

Since it was difficult to change carrier density widely in $n$=4:Hg1234~\cite{Itohara}, we used Ba$_2$Ca$_3$Cu$_4$O$_8$(F$_y$O$_{1-y}$)$_2$ ($n$=4:0234F)~\cite{Iyo1} as $n$=4 compounds. The substitution of O$^{-2}$ for F$^{-1}$ at the apical site increases $p$ and $T_{\rm c}$ from 55 K up to 102 K~\cite{Iyo1,Shimizu2009JPSJ}. As shown in  Figs.~\ref{fig:summary_n4}(a)-\ref{fig:summary_n4}(d), systematic $^{63}$Cu- and $^{19}$F-NMR studies have revealed SC and AFM characteristics for $n$=4:0234F, which are denoted as 0234F($\sharp$1), 0234F($\sharp$2), 0234F($\sharp$3), and 0234F($\sharp$4) with nominal contents $y$= 0.6, 0.7, 0.8 and 1.0, respectively. The measurements of $K_{\rm s}^{ab}$ shown in Figs.~\ref{fig:summary_n4}(e-h) enable us to estimate $p$ at IP and OP for the samples\cite{Shimizu2011PRB}. In fact, $p$ increases progressively with decreasing $y$. 

\begin{figure}[tbp]
\begin{center}
\includegraphics[width=7.5cm]{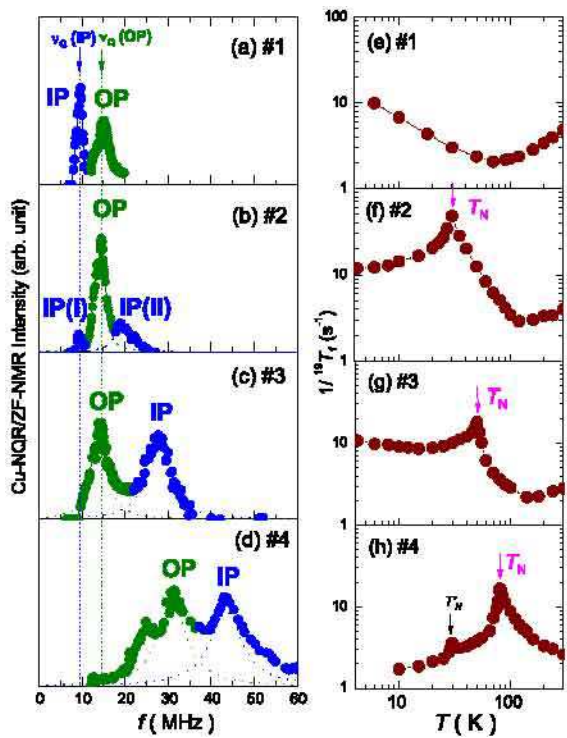}
\end{center}
\caption{\footnotesize (Color online) Cu-NQR/ZFNMR spectra at $B_{\rm ext}$=0 T and $T$=1.5 K for $n$=4:0234F: (a) $\sharp1$, (b) $\sharp2$, (c) $\sharp3$, and (d) $\sharp4$. 
The $T$ dependences of $^{19}$F-NMR ($1/T_1$)s for (e) $\sharp1$, (f) $\sharp2$, (g) $\sharp3$, and (h) $\sharp4$ at $f$=174.2 MHz and $B \parallel$ $c$-axis. The AFM moments and $T_N$s for OPs and IPs are evaluated from these results (see text). [cited from ref.\cite{Shimizu2009JPSJ}] 
}
\label{fig:zero} 
\end{figure}

Figures~\ref{fig:zero}(a)-\ref{fig:zero}(d) show the Cu-NQR/ZF-NMR spectra at $T$=1.5 K for $n$=4:0234F: (a) $\sharp1(y\sim$0.6), (b) $\sharp2(y\sim$0.7), (c) $\sharp3(y\sim$0.8), and (d) $\sharp4(y\sim$1.0). 
For $\sharp1$, two NQR spectra revealed that $^{63}\nu_{\rm Q}$(IP)=9.7 MHz and $^{63}\nu_{\rm Q}$(OP)=15 MHz, which are comparable to those for other paramagnetic multilayered cuprates\cite{Julien,MagishiPRB,Itohara}. For $\sharp2$, the NQR spectrum at OP is observed at $^{63}\nu_{\rm Q}$(OP)=15 MHz.  By contrast, it is not expected that the NQR spectral intensity at IP(i) observed at 9.1 MHz will be significantly smaller than the intensity of the ZF-NMR spectrum at IP(ii). The latter spectrum probes an internal field $B_{\rm int}$=1.5 T in association with the onset of AFM order with $M_{\rm AFM}$(IP)$\sim$ 0.08 $\mu_{\rm B}$. These results point to that the IPs for $\sharp2$ undergo a phase separation into the paramagnetic and AFM phases owing to closeness to $p_c$, at which an AFM order collapses.

For $\sharp3$, the observation of the NQR spectrum at OP indicates that a spontaneous AFM moment is absent, whereas the  NMR spectrum at approximately 28 MHz at IP probes $B_{\rm int}\sim$ 2.4 T and hence $M_{\rm AFM}$(IP)$\sim0.12$ $\mu_{\rm B}$. 
Figure~\ref{fig:zero}(d) shows that the Cu-ZF-NMR spectra at IP and OP are observed at approximately 45 and 30 MHz for $\sharp4$ with $T_{\rm c}$=55 K, which allows us to estimate $B_{\rm int}\sim$ 3.8 T and 2.7 T and hence $M_{\rm AFM}$(IP)$\sim$ 0.18$\mu_{\rm B}$ and $M_{\rm AFM}$(OP)$\sim$ 0.11$\mu_{\rm B}$, respectively. 
Note that no trace of NQR spectra excludes the possibility of phase separation into the paramagnetic and AFM phases. 
Therefore, the OP, which is mainly responsible for the HTSC with $T_{\rm c}$= 55 K, manifests the AFM order, indicating that the uniform mixing of AFM with the spontaneous moment $M_{\rm AFM}$=$0.11 \mu_{\rm B}$ and HTSC at $T_{\rm c}$= 55 K occurs in the OP.  
Taking into account of the ARPES experiment on this compound ($\sharp4$) that observed SC gaps on Fermi sheets of both IP and OP\cite{YChen}, we deduce that the AFM order coexists uniformly with HTSC at both IP and OP for $\sharp4$\cite{KitaokaIOP}.

The $T_{\rm N}$s of $\sharp2$, $\sharp3$, and $\sharp4$ are determined by the $^{19}$F-NMR $1/T_1$ measurements at the $B \parallel c$-axis, as presented for all the samples in Figs.~\ref{fig:zero}(e-h). 
In the present case, $^{19}(1/T_1)$ is dominated by magnetic fluctuations~\cite{Shimizu2009JPSJ}.
Figures~\ref{fig:zero}(f) and \ref{fig:zero}(g) show the $T$ dependences of the $^{19}(1/T_{1})$s of $\sharp2$ and $\sharp3$, which exhibit peaks at approximately 30 and 50 K, respectively. These peaks are associated with the onset of the AFM order at $T_{\rm N}$=30 K at IP(ii) for $\sharp2$ and $T_{\rm N}$=50 K at IP for $\sharp3$, accompanying $M_{\rm AFM}$(IP(ii))$\sim$0.08 $\mu_{\rm B}$ and $M_{\rm AFM}$(IP)$\sim$0.12 $\mu_{\rm B}$, respectively. It is unexpected for the $1/T_1$ of $\sharp4$ to show two peaks at $T_{\rm N} \sim$ 80 K and $T'_{\rm N}\sim$ 30 K, as shown in Fig.~\ref{fig:zero}(h). Since $p$(IP) $<$ $p$(OP) and $M_{\rm AFM}$(IP) $>$ $M_{\rm AFM}$(OP), it is likely that  $T_{\rm N}$s inherent to IP  become larger than that at OP. 
The AFM order inherent to OP may develop below $T'_{\rm N}\sim$ 30 K. Here, note that $T'_{\rm N}\sim$ 30 K at OP for $\sharp4$ is comparable to $T_{\rm N}\sim$ 30 K at IP (ii) for $\sharp2$ because their $p$s are almost the same for both compounds. 

\begin{figure}[tbp]
\begin{center}
\includegraphics[width=7.5cm]{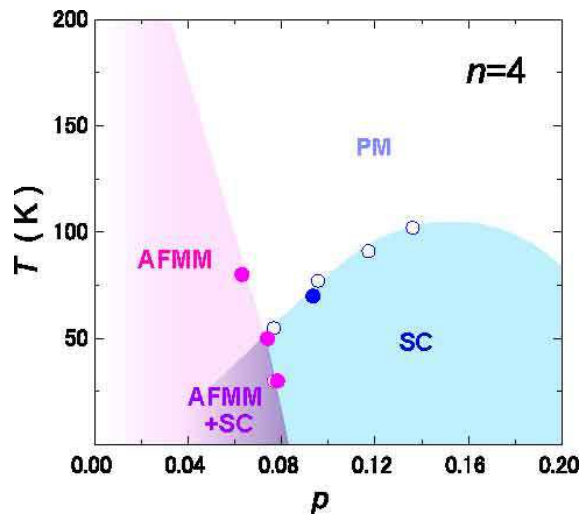}
\end{center}
\caption{\footnotesize (Color online)  Phase diagram of AFM and HTSC at homogeneously doped CuO$_2$ plane in $n$=4 compounds.  $T_{\rm N}$ and  $T_{\rm c}$ are plotted against the hole density $p$.
The solid and empty circles correspond to the data for IP and OP, respectively. 
Note that the uniform mixing of AFM with $T_{\rm N}$=30 K and HTSC with $T_{\rm c}$=55 K takes place at OP of $\sharp4$. 
A critical hole density for the AFM order in $n$=4 compounds is extrapolated to $p_c(4)\sim$ 0.08, which is lower than that for the $n$=5 compounds. The pseudogap phase of the $n$=4 compounds has not been determined yet. [cited from refs.\cite{Shimizu2009JPSJ,Shimizu2011PRB}]
}
\label{fig:PD}
\end{figure}

Figure~\ref{fig:PD} shows the phase diagram of AFM and HTSC for $n$=4:0234F. Here, $T_{\rm c}$ and $T_{\rm N}$ at OP and IP are plotted against $p$\cite{Shimizu2011PRB}. This phase diagram resembles that for the $n$=5 compounds.  In particular, the uniformly mixed phase of AFM with $T_{\rm N}$=30 K and HTSC with $T_{\rm c}$=55 K was observed at the OP for $\sharp4$,  demonstrating that it is the universal phenomenon inherent in a single CuO$_2$ plane in the underdoped regime. 
In the phase diagram of the $n$=4 compounds presented in Fig.~\ref{fig:PD}, $p_c(4)$ at which the AFM order collapses is extrapolated to 0.08, which is smaller than $p_c(5)\simeq$0.1 for the $n$=5 compounds. As discussed in $\S$ 4.1, this is because the interlayer magnetic coupling becomes weaker as the number $n$ of CuO$_2$ layers decreases.

Here, we comment on 0234F($\sharp4$) that attracted much attention at the initial stage of studies~\cite{YChen,OK,Shimizu2007}, since the {\it self-doping} mechanism had been proposed: charge carriers are transferred between IP and OP in a unit cell. Here a formal Cu valence was assumed to be just 2+ in an ideal case of a nominal fluorine content   (F$_2$), where the apical sites are fully occupied by F$^{-1}$. However, extensive investigations on 02($n$-1)$n$F with $n$=2, 3, and 4 have ruled out the possibility of a {\it self-doping} mechanism in these compounds~\cite{Shimizu2009PRB}.
The point was that an inevitable deviation from the nominal apical-fluorine F$^{-1}$ content results in the doping of hole carriers into both OP and IP.

\subsection{Phase diagram of $n$=3 compounds}

\begin{figure}[tbp]
\begin{center}
\includegraphics[width=7.5cm]{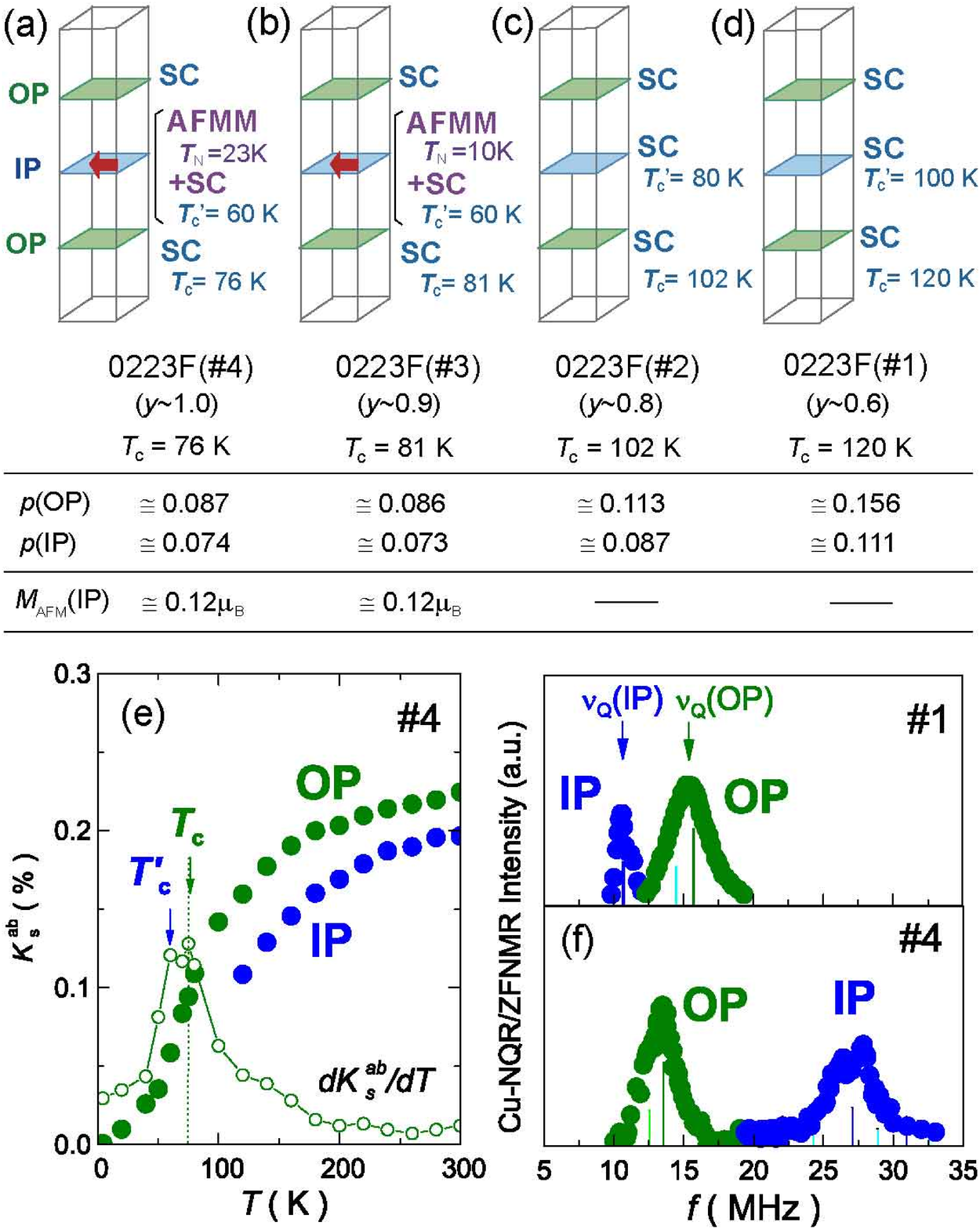}
\end{center}
\caption{\footnotesize (Color online) Illustration of layer-dependent physical properties for n=3:0223F: (a) $\sharp4$($y\sim$1.0), (b) $\sharp3$($y\sim$0.9), (c) $\sharp2$($y\sim$0.8), and (d) $\sharp1$($y\sim$0.6). (e) $T$ dependences of $K_{\rm s}^{ab}$ and $dK_{\rm s}^{ab}/dT$ for $\sharp4$. (f) Cu-NQR/ZFNMR spectrum for $\sharp4$ at 1.5 K, along with that for $\sharp1$ with no AFM order, revealing $M_{\rm AFM}$(IP)$\sim$ 0.12 $\mu_{\rm B}$ and  $M_{\rm AFM}$(OP)=0 for $\sharp4$. 
The middle panels present tables of the hole densities of $p$(IP) and $p$(OP), and the AFM ordered moments at IPs (see text).[cited from refs.\cite{Shimizu2009PRB,Shimizu2011_n3}]
}
\label{fig:summary_n3}
\end{figure}

\begin{figure}[tbp]
\begin{center}
\includegraphics[width=7.5cm]{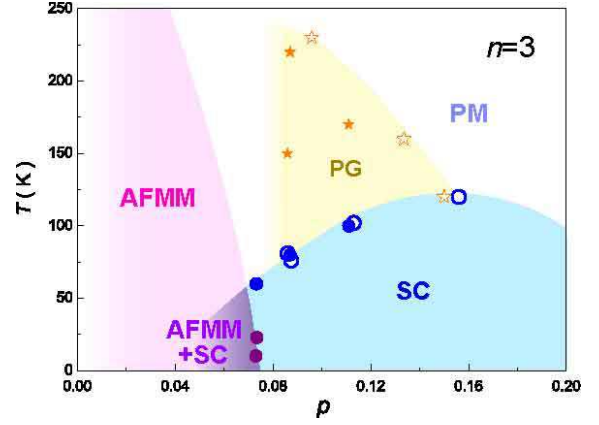}
\end{center}
\caption{\footnotesize (Color online) Phase diagram of $n$=3 compounds. The solid and empty circles correspond to the data for IP and OP, respectively. The critical hole density of AFM order for the $n$=3 compounds is extrapolated to $p_c(3)\sim$ 0.075, which is lower than that for the $n$=4 and $n$=5 compounds. 
The filled and open stars represent the pseudogap temperature $T^{*}$ determined from the peak of $^{63}$Cu-$(1/T_1T)$ for n=3:0223F\cite{Shimizu2011_n3} and $M$1223\cite{Julien,Kotegawa2002,Shimizu2011_n3}, respectively. This suggests that the {\it spin gap} collapses in the AFM-HTSC mixed state for $p < p_c(3)$.
[cited from refs.\cite{Shimizu2009PRB,Shimizu2011_n3}]
}
\label{fig:PD_n3}
\end{figure}

The $^{63}$Cu- and $^{19}$F-NMR studies of three-layered Ba$_2$Ca$_2$Cu$_3$O$_6$(F$_y$O$_{1-y}$)$_2$ compounds with a nominal F content $y$ denoted as $n$=3:0223F have revealed the SC and AFM characteristics shown in Figs.~\ref{fig:summary_n3}(a)-\ref{fig:summary_n3}(d)~\cite{Shimizu2009PRB,Shimizu2011_n3}. 
The $p$s at IP and OP determined by the relationship of $K_{\rm s}^{ab}$(RT) vs $p$ are listed in the middle panel of Fig.~\ref{fig:summary_n3}. 
Figure \ref{fig:summary_n3}(e) shows the $T$ dependences of $K_{\rm s}^{ab}$ and $dK_{\rm s}^{ab}/dT$ for $\sharp4$. 
Besides the bulk $T_{\rm c}$=76 K, $T_{\rm c}'$ inherent in IP is tentatively deduced as $T_{\rm c}'$=60 K from a secondary peak in the $T$ dependence in $dK_{\rm s}^{ab}/dT$. 
Figure~\ref{fig:summary_n3}(f) shows the Cu-NQR/ZFNMR spectrum of $\sharp4$, along with that of $\sharp1$;  the spectra are observed at each NQR frequency, revealing no AFM order at either layer of $\sharp1$. 
In $\sharp4$, the spectrum observed at 13.6 MHz corresponds to the $\nu_{Q}$ at OP, which is slightly lower than that at OP of $\sharp1$ because of the lower doping level\cite{Ohsugi,Zheng,Haase}.  
On the other hand, the spectrum of IP is observed at a frequency much higher than the NQR frequency, which probes $B_{\rm int}$ $\sim$ 2.4 T and hence $M_{\rm AFM}$(IP)$\sim$ 0.12 $\mu_{\rm B}$. 
We note that a phase separation into the AFM and paramagnetic phases is excluded in the IP of $\sharp4$, because no paramagnetic NQR spectrum for IP was observed. 
The $T_{\rm N}$=23 K at IP was determined from the peak in  $^{19}$F-NMR $1/T_1$ (see ref.\cite{Shimizu2009PRB}). 
Details of systematic NMR experiments on $\sharp1$, $\sharp2$, and $\sharp3$ were published elsewhere \cite{Shimizu2011_n3}. 
As a consequence, we have unraveled a phase diagram of the $n$=3:0223F compounds\cite{Shimizu2011_n3}, as shown in Fig. \ref{fig:PD_n3}. 
The AFM order emerges at IPs of $\sharp3$ and  $\sharp4$ with $p\le$ 0.075, suggesting that the critical hole density of AFM order for $n$=3 compounds $p_c(3)$ is close to 0.075, which is lower than $p_c(4)\simeq$0.08 for the $n$=4 compounds and $p_c(5)\simeq$0.10 for the $n$=5 compounds. 
These AFM ordered states at IPs appear under the background of HTSC with $T_{\rm c}'$= 60 K, which indicates that the uniformly mixed state of AFM and HTSC emerges universally for a single CuO$_2$ layer in the underdoped region, when a magnetic interlayer coupling is strong enough to stabilize an AFM order.

\begin{figure}[tbp]
\begin{center}
\includegraphics[width=7.5cm]{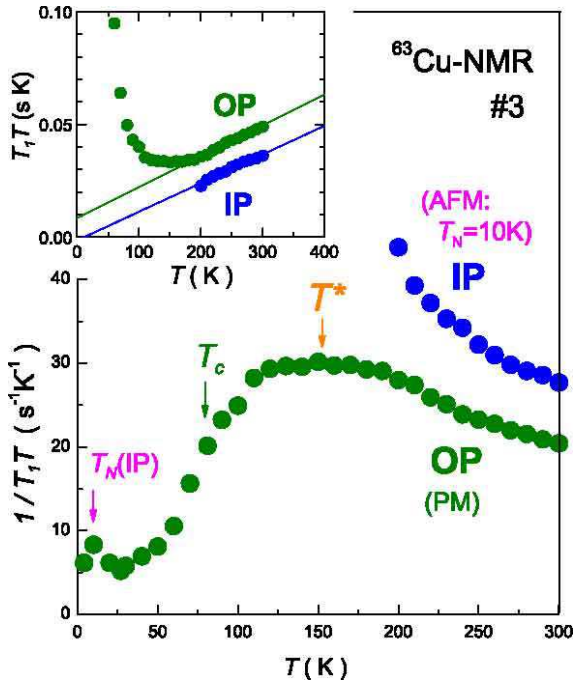}
\end{center}
\caption{\footnotesize (Color online) $T$ dependences of $^{63}$Cu-NMR 1/$T_1T$ at  OP and IP for $\sharp$3 at $B\perp c$ and $f$=174.2 MHz\cite{Shimizu2011_n3}.  $1/T_1T$ at IP  continues to increase upon cooling towards the AFM order at $T_{\rm N}\sim$ 10 K, whereas $1/T_1T$ at OP exhibits a pseudogap behavior with $T^* \sim$150 K.  The $p$ dependence of $T^*$ for $n$=3 compounds is shown in Fig. \ref{fig:PD_n3}. 
The inset shows a plot of $T_1T$ vs $T$ to reveal the Curie-Weiss behavior at high temperatures as $1/T_1T$=$C/(T+\theta)$. As for IP, $\theta$=-10 K was obtained, which corresponds to the peak of $1/T_1T$ at OP, that is, $T_N\sim$10 K at IP \cite{Shimizu2011_n3}. 
[cited from ref.\cite{Shimizu2011_n3}]
}
\label{fig:n3_PG}
\end{figure}

To probe the pseudogap behavior for the $n$=3 compounds, the $^{63}$Cu-NMR $1/T_1T$ measurement has been  performed in $\sharp$1, $\sharp$2, and $\sharp$3 \cite{Shimizu2011_n3}. 
Figure \ref{fig:n3_PG} shows the $T$ dependence of $1/T_1T$ for $\sharp$3, which starts to decrease upon cooling below $T^* \sim$ 150 K for OP. On the other hand, the $1/T_1T$ at IP in the same sample continues to increase upon cooling down to 200 K, and the NMR spectrum at IP is lost below 200 K because of extremely short nuclear relaxation times due to the critical enhancement of AFM spin fluctuations. 
Taking into account of the doping level ($p\sim0.073$), unless an AFM order occurs, one could expect $T^*$ to be around 200 $\sim$ 250 K, which is deduced from the case of the single- and bilayered cuprates presented in Fig. \ref{fig:PG}, but no peak of $1/T_1T$ was seen above 200 K. 
Instead, the inset of Fig. \ref{fig:n3_PG} indicates that the $1/T_1T$ for IP at high temperatures follows the Curie-Weiss law as $1/T_1T$=$C/(T+\theta)$ with $\theta \sim -$ 10 K. 
This indicates that the {\it out-of-plane} magnetic interaction, which is responsible for the onset of the AFM order at $T_{\rm N}\sim$ 10 K, causes the {\it in-plane} AFM correlation to develop further without opening a gap at the low-energy spectral weight of AFM excitations.  
Although  $1/T_1T$  is expected to decrease owing to the opening of the SC gap, we consider that the low-energy parts in AFM excitations  below $T_c$  continue to increase at IP down to $T_{\rm N}\sim$ 10 K, since the peak at  $T\sim 10$ K observed in the $1/T_1T$ for OP points to some divergence of the $1/T_1T$ at IP towards the AFM order at $T_{\rm N}\sim$ 10 K through  the supertransferred hyperfine coupling between $^{63}$Cu nuclei at OP and the Cu-derived spin at IP.  When noting that $\theta$ coincides with $T_{\rm N}$, as  far as the low-energy parts in AFM excitations are concerned in an energy region smaller than an SC energy gap, we suppose that a {\it spin gap} may not open between 10 K and 200 K.

\subsection{AFM and HTSC in $n$=8 compound}

To gain further insight into the $n$ dependence of $p_c(n)$ at which the AFM order collapses, we have investigated the  magnetic and SC properties of $n$=8:Hg1278(OPT) with $T_{\rm c}$=103 K~\cite{Yamaguchi}. Here, note that the interlayer magnetic coupling is expected to be larger than in the other multilayered cuprates. 
As indicated in Figs.~\ref{fig:Tc_n}(a) and \ref{fig:Tc_n}(b), note that the $c$-axis length increases progressively with $n$, whereas $T_{\rm c}$ is almost constant for $n > 6$. The sample for NMR study comprises an almost single phase of $n$=8:Hg1278, although it is inevitable that a small number of stacking faults along the $c$-axis are contained for layered cuprates with $n > 6$~\cite{Iyo_TcVsn,Iyo_Hg_F}. 

\begin{figure}[tbp]
\begin{center}
\includegraphics[width=7.5cm]{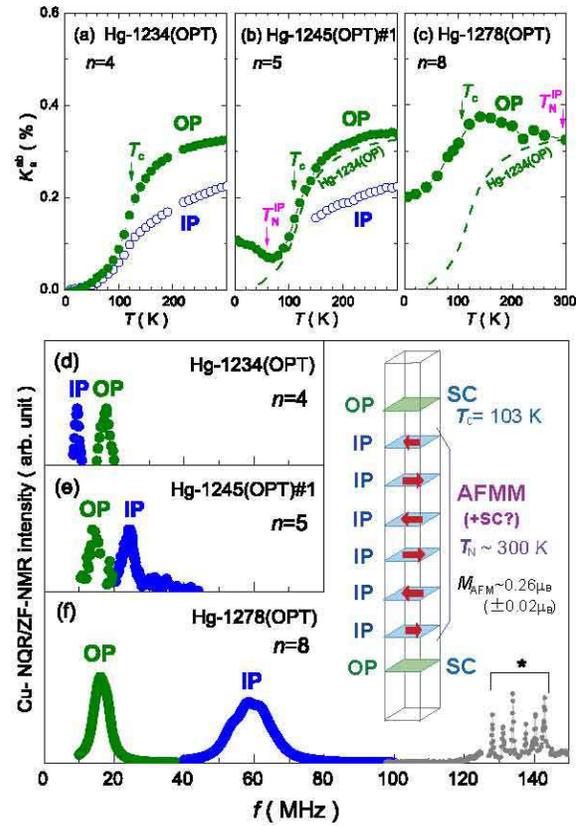}
\end{center}
\caption{\footnotesize (Color online)  $T$ dependence of $K_{\rm s}^{ab}$ at OP in nearly optimally doped Hg12($n$-1)$n$ series: (a) $n$=4:Hg1234(OPT) with $T_{\rm c}$=123 K\cite{Kotegawa2001}, (b) $n$=5:Hg1245(OPT)$\sharp$1 with $T_{\rm c}$=108 K\cite{Kotegawa2004}, and (c) $n$=8:Hg1278(OPT) with $T_{\rm c}$=103 K.  
The broken curves in (b) and (c) show the $T$ dependence of  $K_{\rm s}^{ab}$ at OP of Hg1234(OPT), which shows no AFM order.
The Cu-NQR/ZFNMR spectra at $B_{\rm ext}$=0 and $T$=1.5 K for (d) $n$=4:Hg1234(OPT), (e) $n$=5:Hg1245(OPT)$\sharp$1\cite{Kotegawa2004}, and (f) $n$=8:Hg1278(OPT). The spectra indicated by the symbol ($\ast$) arise from the AFM-Mott insulator CaCuO$_2$(IL), which is contaminated during the high-pressure synthesis process~\cite{Takano}.
The inset at the bottom illustrates the physical properties of $n$=8:Hg1278(OPT) with $T_{\rm c}$=103 K. [cited from ref.\cite{Yamaguchi}]
}
\label{fig:n8}
\end{figure}

Figures~\ref{fig:n8}(a)-\ref{fig:n8}(c) show the $T$ dependence of $K_{\rm s}^{ab}$ at OP of (c) $n$=8:Hg1278(OPT), together with (a) $n$=4:Hg1234(OPT) and (b) $n$=5:Hg1245(OPT)$\sharp$1, which are as-prepared samples in an optimally doped regime. 
$K_{\rm s}^{ab}$(RT) at OP is comparable to those for $n$=4:Hg1234(OPT) and $n$=5:Hg1245(OPT)$\sharp$1, pointing to $p($OP$)\sim$ 0.14 in these compounds. 
It is also corroborated by similar NQR frequencies of OP in these compounds as will be discussed later. 
Therefore, if all layers were in the paramagnetic state, the $T$ dependence of $K_{\rm s}^{ab}$ should be similar to that of $n$=4:Hg1234(OPT) as indicated by the broken curves in Figs.~\ref{fig:n8}(b) and \ref{fig:n8}(c). 
By contrast, note that the $K_{\rm s}^{ab}$ at OP for $n$=5:Hg1245(OPT)$\sharp$1 deviates from this broken curve below $T_{\rm N}$=60 K at which the AFM order sets in at IPs. 
Accordingly, the $T_{\rm N}$ of $n$=8:Hg1278(OPT) is tentatively deduced to be about 300 K, since $K_{\rm s}^{ab}$ deviates from the broken curve below 300 K, as shown in Fig.~\ref{fig:n8}(c).
Although $n$=8:Hg1278(OPT) comprises three crystallographically inequivalent IPs, no Cu-NMR spectra at IPs were observed up to 300 K, which suggests that AFM order emerges at all IPs. 
To evaluate AFM moments of IPs, we present the Cu-NQR/ZF-NMR spectra of Hg1278(OPT) at $T$=1.5 K in Fig.~\ref{fig:n8}(f), which are compared with the spectra of (d) $n$=4:Hg1234(OPT) and (e) $n$=5:Hg1245(OPT)$\sharp$1. Since $\nu_Q$(OP)$\sim$ 16 MHz at OP for $n$=8:Hg1278(OPT) coincides with those for $n$=4 and $n$=5 compounds, the OP is paramagnetic. 
The Cu-ZF-NMR spectrum at approximately 50$\sim$70 MHz can be assigned to IPs, although we cannot resolve three inequivalent IPs precisely from the present ZF-NMR spectra. 
The $M_{\rm AFM}$ at those IPs may be tentatively estimated as $M_{\rm AFM}$(IPs)=0.24 $\sim$ 0.28$\mu_{\rm B}$. 
The ZF-NMR spectra at approximately 120 $\sim$ 150 MHz arise from the AFM-Mott insulator CaCuO$_2$(IL), which is contaminated during synthesis process under high-pressure and high-temperature conditions~\cite{Takano}.
Further extensive NMR measurements are desired for cuprates with more than six layers over a wide doping level. 

Nevertheless, note that $T_{\rm c}$=103 K for $n$=8:Hg1278(OPT) remains high, even though the OPs that are responsible for HTSC are separated by a thick AFM block consisting of six IPs. 
When noting that SC Cooper pairs can tunnel between OPs through AFM ordered IPs by virtue of the Josephson coupling as predicted by a theoretical study of multilayered cuprates~\cite{Mori}, it would be expected that HTSC can maintain relatively high and constant $T_{\rm c}$ values, as shown in Fig.~\ref{fig:Tc_n}(b). 

More importantly, when noting that the OP with $p$(OP)$\sim$ 0.14 is paramagnetic, we remark that the critical hole density of $n$=8 compounds $p_c(8)$ does not reach 0.14 even when an interlayer magnetic coupling is sufficiently increased. 
This result is consistent with the fact that $M_{\rm AFM}(p)$ at 1.5 K decreases monotonically as $p$ increases and is extrapolated to zero in the range of $p_c(M_{\rm AFM}$=$0) \le $ 0.14, as discussed in $\S$ 4.2 (see Fig.~\ref{Mvsp}).

\section{Discussion}

\subsection{Number $n$ of CuO$_2$ layer dependence of phase diagram of AFM and HTSC}

\begin{figure*}[t]
\centering
\includegraphics[width=12cm]{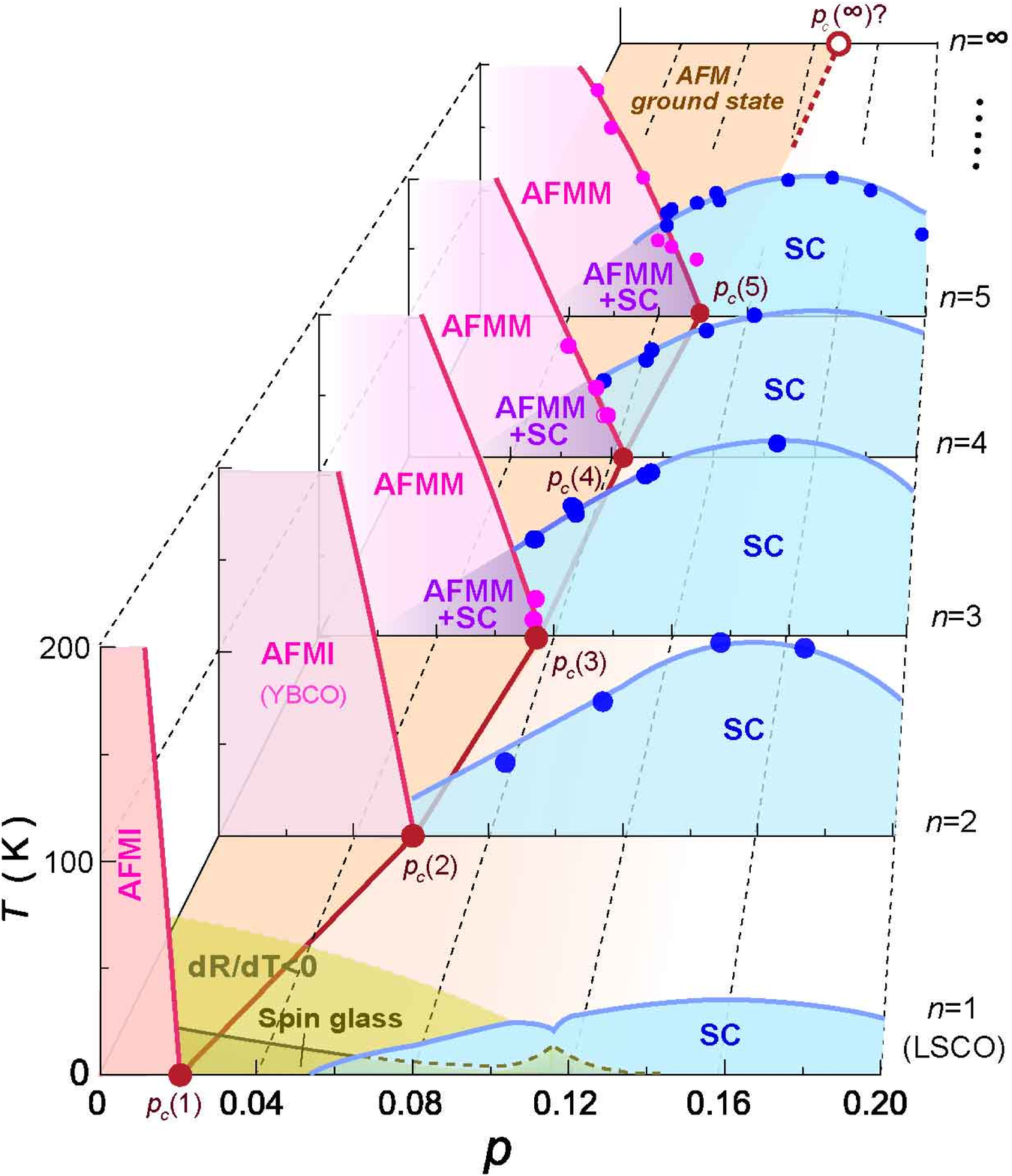}
\caption[]{\footnotesize (Color online) Phase diagrams of AFM and HTSC for $n$=1:LSCO~\cite{Keimer,JulienSG}, $n$=2:0212F~\cite{Shimizu2011PRB} and YBCO$_{6+x}$~\cite{Sanna,Coneri}, $n$=3:0223F~\cite{Shimizu2009PRB,Shimizu2011_n3}, $n$=4:0234F~\cite{Shimizu2007,Shimizu2009PRB,Shimizu2009JPSJ}, and $n$=5:$M$1245~\cite{Kotegawa2004,MukudaPRL2006,Mukuda2008,Mukuda2010,Tabata}.  $p_{c}(n)$, at which an AFM order collapses, increases from $p_{c}(n)\sim$0.075 to 0.08 to 0.10 as $n$ increases from 3 to 4 to 5, respectively. The result on $n$=8 compound has revealed that $p_c(8)$ does not reach 0.14 even when an interlayer magnetic coupling is enhanced, which is also deduced from the fact that $M_{\rm AFM}$ decreases to zero in the range of $p_c(M_{\rm AFM}$=$0) \le $ 0.14 (see Fig. \ref{Mvsp}).
[cited from refs.\cite{Mukuda2010,KitaokaJPCS2011,KitaokaIOP}]
}
\label{fig:PhaseDiagram}
\end{figure*}

Figure~\ref{fig:PhaseDiagram} shows the phase diagrams of AFM and HTSC for $n$=1: LSCO~\cite{Keimer,JulienSG}, $n$=2: 0212F\cite{Shimizu2011PRB} and YBCO$_{6+x}$~\cite{Sanna,Coneri}, $n$=3:0223F~\cite{Shimizu2009PRB,Shimizu2011_n3}, $n$=4:0234F~\cite{Shimizu2007,Shimizu2009PRB,Shimizu2009JPSJ}, and $n$=5:$M$1245~\cite{Kotegawa2004,MukudaPRL2006,Mukuda2008,Mukuda2010,Tabata}.  
The phase diagram on $n$-layered cuprates with $n$=3, 4, and 5 are characterized as follows: 
(i) The AFM metallic phase is robust up to an optimally hole-doped region: The quantum critical point ($p_{c}$) at which an AFM order collapses increases from $p_{c}(n)\sim$0.075 to 0.08 to 0.10 as $n$ increases from 3 to 4 to 5, respectively.
(ii) The uniform coexistence of AFM and HTSC is the universal phenomenon inherent in a single CuO$_2$ plane in an underdoped region.
(iii) The maximum $T_{\rm c}$ takes place at approximately $p(T_{\rm c}^{\rm max})\sim$~0.16 irrespective of $n$. 

\begin{figure}[tbp]
\centering
\includegraphics[width=7.5cm]{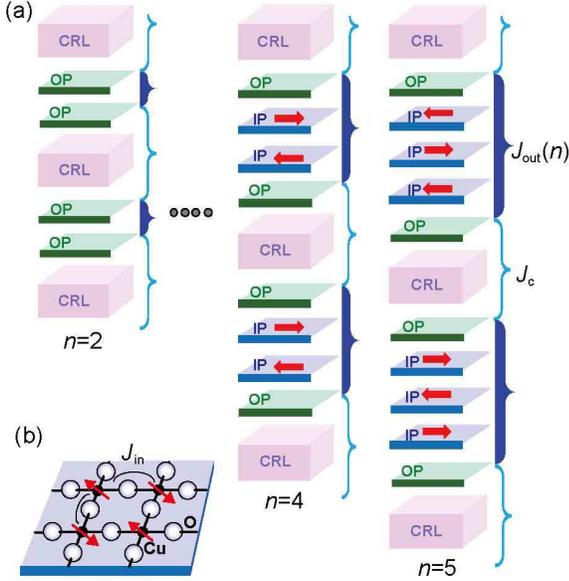}
\caption[]{\footnotesize (Color online) Schematic illustrations  of magnetic couplings in $n$-layered cuprates: (a) interlayer magnetic coupling along $c$-axis and (b) in-plane superexchange interaction $J_{\rm in}$  among spins at nearest-neighbor Cu sites in two-dimensional lattice of CuO$_2$ plane. Here, $J_{\rm c}$ is the magnetic coupling between OPs through CRL, which is independent of $n$, and $J_{\rm out}(n)$ is the magnetic coupling between OPs through IPs, which increases with $n$. Since $J_{\rm in}$ is as large as 1300~K in undoped AFM-Mott insulators\cite{JinLSCO,JinYBCO1,JinYBCO2,TokuraJ}, an AFM order is stabilized when the interlayer magnetic coupling ($\sqrt{J_cJ_{\rm out}(n)}$) becomes stronger with increasing $n$. 
}
\label{fig:interlayer}
\end{figure}

The phase diagrams of $n$-layered cuprates with $n$=3, 4, and 5 differ significantly from the well-established phase diagrams of $n$=1:LSCO and $n$=2:YBCO$_{6+x}$,  where $p_c(1)\sim$ 0.02~\cite{Keimer} and $p_c(2)\sim$ 0.055~\cite{Sanna,Coneri}, respectively, as shown in Fig.~\ref{fig:PhaseDiagram}. 
In fact, from the present NMR measurement, the phase diagram of $n$=2:0212F does not reveal an AFM order in the range of $p\ge 0.083$.\cite{Shimizu2011PRB} It is apparent that as $n$ increases from 1 to 5, $p_c(n)$ increases from 0.02 to 0.10. 
The mother compounds for HTSC are characterized by a large in-plane superexchange interaction $J_{\rm in}\sim$1300 K among spins at nearest-neighbor Cu sites~\cite{JinLSCO,JinYBCO1,JinYBCO2,TokuraJ}.
However, since no long-range AFM order occurs at a finite temperature for an isolated two-dimensional (2D) system, the interlayer magnetic coupling along the $c$-axis, which depends on $n$, plays a crucial role in stabilizing an AFM order. 
An effective interlayer magnetic coupling of $n$-layered cuprates is given as $\sqrt{J_cJ_{\rm out}(n)}$, where $J_{\rm c}$ is the magnetic coupling between OPs through CRL and $J_{\rm out}(n)$ is the magnetic coupling in a unit cell, as illustrated in Fig. \ref{fig:interlayer}(a). 
Here, $J_{\rm c}$ is independent of $n$, but $J_{\rm out}(n)$ increases with increasing $n$. 
In this context, it is the weak interlayer magnetic coupling that suppresses the static long-range AFM order in LSCO and YBCO$_{6+x}$ at such small carrier densities. 
In the contrast, the result for the $n$=8 compound has revealed that $p_c(8)$ does not reach 0.14 even when the interlayer magnetic coupling becomes sufficiently large. 
It is likely that $p_c$ saturates at approximately 0.14 even in the strong limit of interlayer magnetic coupling expected for the $n$=$\infty$ compound, since $M_{\rm AFM}$ in the ground state is extrapolated to zero at $p_c(M_{\rm AFM}$=$0)\le$ 0.14, as discussed in the next section (see Fig. \ref{Mvsp}).

\subsection{Ground-state phase diagram of AFM  and HTSC}

\begin{figure*}[t]
\centering
\includegraphics[width=14cm]{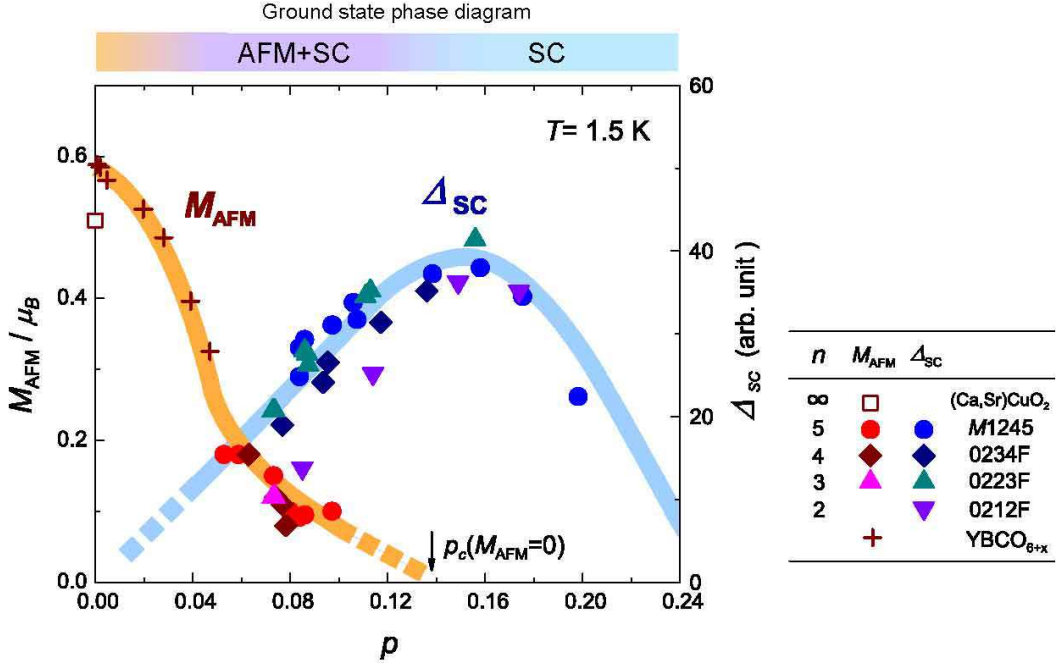}
\caption[]{\footnotesize (Color online) Ground-state phase diagram: $p$-dependence of $M_{\rm AFM}$ at $T$=1.5 K and SC gap ($\Delta_{\rm SC}$) for $n$=2:0212F\cite{Shimizu2011PRB}, $n$=3:0223F\cite{Shimizu2009PRB}, $n$=4:0234F\cite{Shimizu2009JPSJ}, and $n$=5:$M$1245\cite{Mukuda2008,Mukuda2010,Tabata}. Data of $M_{\rm AFM}$s for nondoped and slightly doped Mott insulating states are cited from $n$=$\infty$:Ca$_{0.85}$Sr$_{0.15}$CuO$_{2}$\cite{Vaknin1989} and $n$=2:YBCO$_{6+x}$\cite{Coneri}, respectively. 
The AFM moment at the CuO$_2$ plane totally disappears in the ground state when $p$= 0.12$\sim$0.14 ($\equiv p_c(M_{\rm AFM}$=0)), which is extrapolated from the $p$ dependence of $M_{\rm AFM}$(solid curve).
$\Delta_{\rm SC}$ shows a maximum just outside of $p_c(M_{\rm AFM}$=0)$\sim$0.14 irrespective of $n$. Here, the $\Delta_{\rm SC}$ is estimated from the $T_{\rm c}$ values with the relation 2$\Delta_{\rm SC}$=$8k_{\rm B}T_{\rm c}$. [cited from refs.\cite{Mukuda2010,KitaokaJPCS2011,KitaokaIOP,Shimizu_n5}]
}
\label{Mvsp}
\end{figure*}

Since $T_{\rm N}$ depends on the strength of interlayer magnetic coupling, i.e., the number of CuO$_2$ layers, the temperature phase diagram against $p$ depends on $n$, as shown in Fig.~\ref{fig:PhaseDiagram}. 
Thus, we discuss here a ground-state phase diagram of a CuO$_2$ plane by plotting the AFM moment~($M_{\rm AFM}$) and SC gap ($\Delta_{\rm SC}$) against $p$, which also gives us an opportunity to compare the experimental outcomes with theoretical ones.

Figure~\ref{Mvsp} shows the $p$ dependence of the $M_{\rm AFM}$ at $T$=1.5~K and $\Delta_{\rm SC}$ evaluated using the relation 2$\Delta_{\rm SC}$=$8k_{\rm B}T_{\rm c}$~\cite{Mukuda2008,Shimizu2009PRB,Shimizu2009JPSJ,Mukuda2010,KitaokaJPCS2011,KitaokaIOP,Shimizu_n5}. 
The $M_{\rm AFM}$ on the CuO$_2$ plane decreases monotonically as $p$ increases and is extrapolated to zero in the range of 0.12$<p_c(M_{\rm AFM}$=$0) \le$0.14, as shown by the solid curve in Fig.~\ref{Mvsp}. 
Here, the data of $M_{\rm AFM}$s for nondoped and slightly doped Mott insulating states are cited from $n$=$\infty$:Ca$_{0.85}$Sr$_{0.15}$CuO$_{2}$\cite{Vaknin1989} and $n$=2:YBCO$_x$\cite{Coneri}, respectively. 
Figures \ref{fig:PhaseDiagram} and \ref{Mvsp} reveal that $p_c(M_{\rm AFM}$=$0)\sim$0.14 is an intrinsic quantum critical hole density at the CuO$_2$ plane, at which the AFM moment totally disappears in the ground state. 
It is noteworthy that the maxima of $\Delta_{\rm SC}$ and $T_{\rm c}$ are at $p(T_{\rm c}^{\rm max})\sim$ 0.16 irrespective of $n$, which are close to $p_c(M_{\rm AFM}$=$0)\sim$0.14, revealing an intimate relationship between the SC order parameter and the AFM ordered moment. 
The phase diagram of $M_{\rm AFM}$ vs $p$ presented here is totally consistent with the ground-state phase diagram at a single CuO$_2$ plane, which has been theoretically addressed thus far in terms of either the $t$-$J$ model~\cite{Chen,Giamarchi,Inaba,Lee1,Himeda,Kotliar,Paramekanti1,Paramekanti2,Lee2,Shih1,Shih2,Yamase1,Yamase2,Ogata,Pathak}, or the Hubbard model in a strong correlation regime~\cite{Senechal,Capone}. Note that $p_c(M_{\rm AFM}$=$0)\sim$ 0.14 determined experimentally is also in good agreement with the theoretical calculation at $T$=0. 

We also point out that no static AFM order with a 3D long-range character can be observed in the range of $p_c(n) < p \le p_c(M_{\rm AFM}$=$0)$ owing to the strong 2D fluctuations and weak interlayer magnetic coupling; hence, the AFM moments at CuO$_2$ planes may fluctuate on a time scale faster than the nanosecond order in NMR measurement, which may be related to various anomalous magnetic behaviors in underdoped regions, as addressed in $\S$ 4.6.

\subsection{Superexchange interaction in hole-doped CuO$_2$ plane}

\begin{figure}[tbp]
\centering
\includegraphics[width=7.5cm]{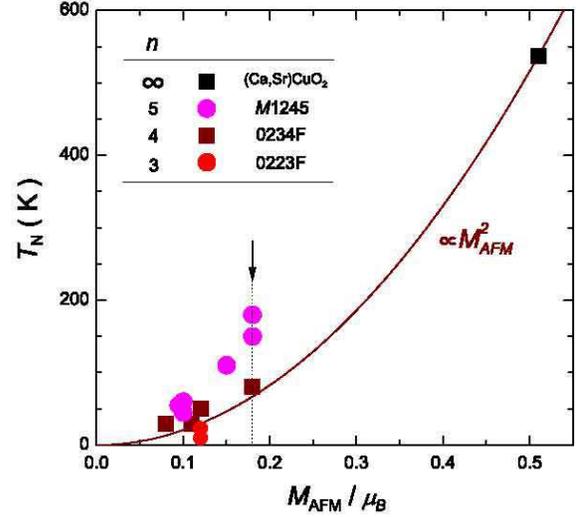}
\caption[]{\footnotesize (Color online) Plot of $T_{\rm N}$ vs $M_{\rm AFM}$~\cite{Shimizu2009PRB,Shimizu2009JPSJ,Mukuda2008,Mukuda2010}. The solid curve shows $T_{\rm N}\propto M_{\rm AFM}^2$ with $T_{\rm N}$=537~K at $M_{\rm AFM}$=0.51$\mu_{\rm B}$ in $n$=$\infty$:Ca$_{0.85}$Sr$_{0.15}$CuO$_{2}$\cite{Vaknin1989}. The arrow points to $M_{\rm AFM}\sim$~0.18$\mu_{\rm B}$.[cited from refs.\cite{Mukuda2010,KitaokaJPCS2011,KitaokaIOP}]
}
\label{MvsTN}
\end{figure}

Figure~\ref{MvsTN} shows a plot of $T_{\rm N}$ vs $M_{\rm AFM}$ to gain insight into the $p$ dependence of the in-plane superexchange interaction $J_{\rm in}(p)$.  In this figure, data are presented with respect to $n$=3:0223F~\cite{Shimizu2009PRB}, $n$=4:0234F~\cite{Shimizu2009JPSJ}, and $n$=5:$M$1245~\cite{Mukuda2008,Mukuda2010}, along with the data of $n$=$\infty$:Ca$_{0.85}$Sr$_{0.15}$CuO$_{2}$ with $p$=0~\cite{Vaknin1989}. 

On the basis of the mean-field approximation of localized spins, we assume that $T_N\propto$~$M_{\rm AFM}^2$, and $J_{\rm out}(n=\infty)$ and $J_{\rm in}(p=0)$ for Ca$_{0.85}$Sr$_{0.15}$CuO$_{2}$ stay constant regardless of $p$. 
The solid curve in Fig.~\ref{MvsTN} shows a graph of $T_N\propto$~$M_{\rm AFM}^2$ plotted using $T_N(\infty)$=537~K at $M_{\rm AFM}$=0.51~$\mu_{\rm B}$ for Ca$_{0.85}$Sr$_{0.15}$CuO$_{2}$. 
First, we notice that, as shown by an arrow in Fig.~\ref{MvsTN}, $T_N(n)$ increases from $n$=4 to 5 as $n$ increases, even though $M_{\rm AFM}\sim$~0.18~$\mu_{\rm B}$ remains constant for the $n$=4 and 5 compounds, which is attributed to the increase in $J_{\rm out}(n)$, namely, $J_{\rm out}(4)<J_{\rm out}(5)$. 
The effective interlayer coupling of the $n$=4 and 5 compounds, given by $\sqrt{J_c J_{\rm out}(n)}$, is always smaller than $J_{\rm out}(\infty)$ in Ca$_{0.85}$Sr$_{0.15}$CuO$_{2}$: However, as shown by the solid curve in Fig.~\ref{MvsTN}, $T_N(n)$s for the $n$=4 and 5 compounds, which are given by $T_N(n)\sim M_{\rm AFM}^2(p)[J_{\rm in}(p)\sqrt{J_cJ_{\rm out}(n)}]^{1/2}$, are larger than $T_N(\infty)$ for $n$=$\infty$:Ca$_{0.85}$Sr$_{0.15}$CuO$_{2}$, which is given by $T_N(\infty)\sim M_{\rm AFM}^2(p)[J_{\rm in}(0)J_{\rm out}(\infty)]^{1/2}$ with $T_N$=537 K at $M_{\rm AFM}$=0.51$\mu_{\rm B}$ and $p$=0. 
Thus, we also obtain an unexpected relation, i.e., $J_{\rm in}(p)>J_{\rm in}(0)\sim$~1300~K. 
The two experimental relationships $-$ the plot of $M_{\rm AFM}$ vs $p$ shown in Fig.~\ref{Mvsp} and the plot of $T_{\rm N}$ vs $M_{\rm AFM}$ shown in Fig.~\ref{MvsTN} $-$ suggest that the AFM ground state in the homogeneously hole-doped CuO$_2$ planes is determined by $p$, $\sqrt{J_cJ_{\rm out}(n)}$, and $J_{\rm in}(p)$ that is larger than $J_{\rm in}(0)\sim$~1300~K. It is surprising that $J_{\rm in}(p)$ becomes stronger for the doped CuO$_2$ planes with AFM order than for the undoped AFM-Mott insulators.  Mean-field theories of HTSC used to consider the superexchange interaction $J_{\rm in}$ as the source of an instantaneous attraction that led to pairing in a $d$-wave state~\cite{Anderson2}. 
The present outcomes may experimentally support such a scenario as far as the underdoped region is concerned, where AFM and HTSC uniformly coexist at a CuO$_2$ plane.
Further theoretical study is desired to address whether $J_{\rm in}$($p$) becomes larger than that for AFM-Mott insulators even though mobile holes are doped. 

\subsection{Pseudo gap behavior in underdoped region }

\begin{figure}[tbp]
\begin{center}
\includegraphics[width=7.5cm]{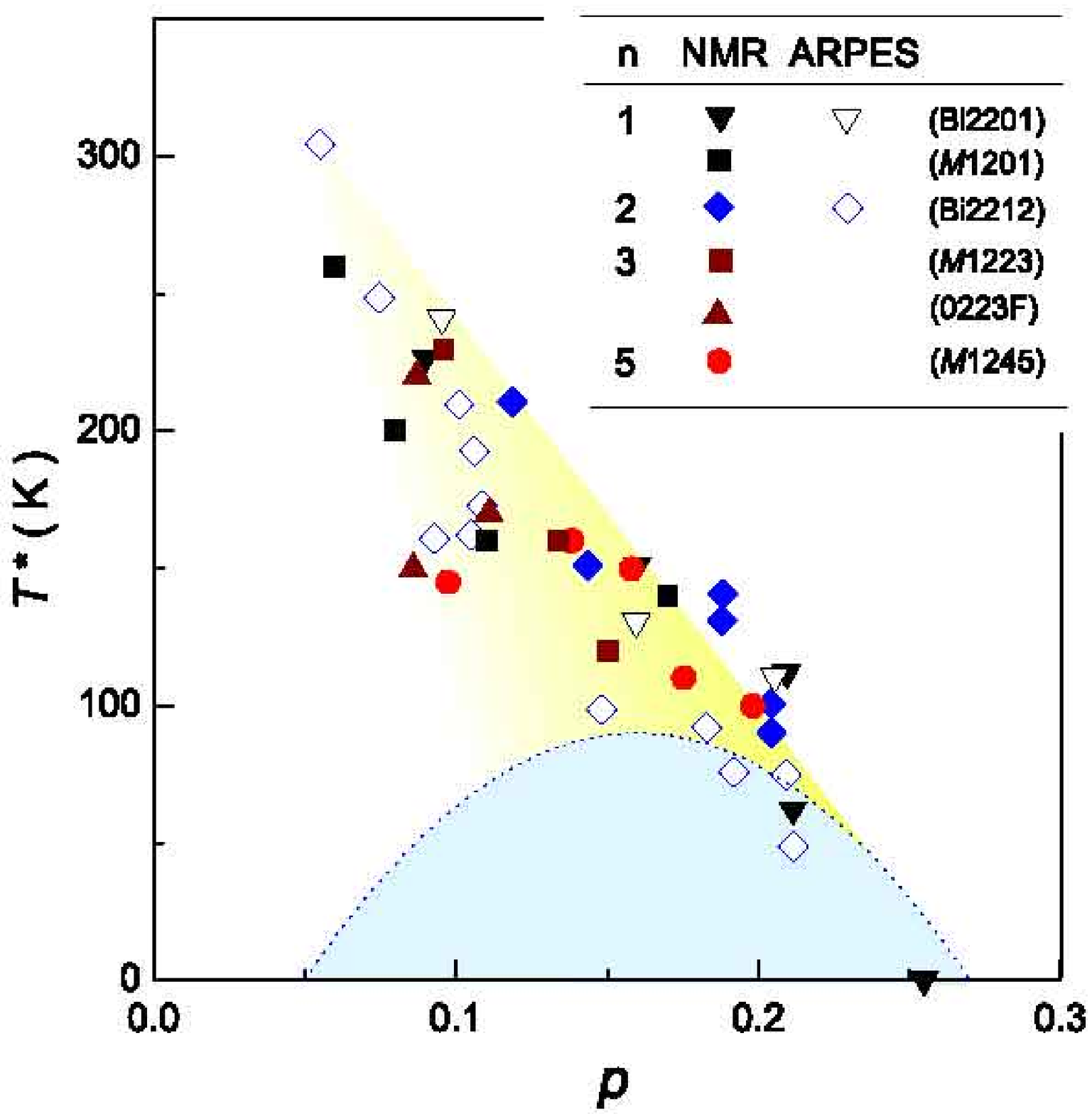}
\end{center}
\caption{\footnotesize (Color online)  $p$ dependences of $T^{*}$ for $n$=5:$M$1245, along with the $T^{*}$ for $n$=1:Bi2201\cite{Kondo,ZhengPG} and Hg1201\cite{ItohHg1201_96}, $n$=2:Bi2212\cite{Campzano,IshidaBi2212,Walstedt}, and $n$=3:0223F\cite{Shimizu2011_n3} and $M$1223\cite{Julien,Kotegawa2002}. Those data were obtained by the measurements of ARPES~\cite{Kondo,Campzano} and NMR~\cite{ZhengPG,ItohHg1201_96,IshidaBi2212,Walstedt,Julien,Kotegawa2002,Shimizu2011_n3}. The broken curve represents the $p$ dependence of $T_{\rm c}$, $T_{\rm c}$=$T_{\rm c}^{\rm max}$[1-82.6($p$-0.16)$^2$].\cite{Groen,Presland}
As far as no AFM order takes place, the $p$ dependences of $T^{*}_{spin}$ and $T^{*}_{charge}$ resemble irrespective of $n$, suggesting that $T^{*}$ is determined by {\it in-plane} magnetic and charge correlations. [cited from ref. \cite{Shimizu2011_n3}]
}
\label{fig:PG}
\end{figure}

The pseudogap behavior emerging above $T_{\rm c}$ is the underlying issue in cuprate superconductors. 
NMR and ARPES studies have observed pseudogap behaviors in spin and charge excitations below $T^{*}_{spin}$ and $T^{*}_{charge}$, respectively. Figure~\ref{fig:PG} shows the $p$ dependence of $T^{*}$ for $n$=5:$M$1245, along with $T^{*}$ for $n$=1:Bi2201~\cite{Kondo,ZhengPG} and Hg1201~\cite{ItohHg1201_96}, $n$=2:Bi2212~\cite{Campzano,IshidaBi2212,Walstedt}, and $n$=3:0223F\cite{Shimizu2011_n3} and $M$1223~\cite{Julien,Kotegawa2002}. 
As shown in Fig.~\ref{fig:PG}, the $p$ dependences of $T^{*}_{spin}$ and $T^{*}_{charge}$ resemble each other for the $n$=1 and $n$=2 compounds, as argued in the literature~\cite{Yoshida}. 
Note that, as far as no AFM order does not takes place, the $p$ dependence of $T^{*}_{spin}$ in the $n$=5 compounds resembles those for the $n$=1, 2, and 3 compounds, suggesting that the magnetic and charge excitations are suppressed below the same $T$, that is, $T^{*}_{spin}$ coincides with $T^{*}_{charge}$. 
Despite the stronger magnetic interlayer coupling in the $n$=5 compounds than in the other multilayered cuprates, the $p$-dependence of $T^{*}_{spin}$ resembles those of $n$=1 and $n$=2 compounds in the optimally doped regime: it is expected that $T^{*}$ is determined by in-plane magnetic and charge correlations~\cite{Yoshida}. From a theoretical point of view, the $t$-$J$ model has explained $T^{*}$ as an onset temperature for a singlet pairing of spinons in the CuO$_2$ plane~\cite{Ogata}, which would be insensitive to the presence of magnetic interlayer coupling. 

Here, we highlight that the spin-gap behavior disappears when the AFM order takes place in $p < p_c(n)$; $1/T_1T$ continues to increase toward $T_{\rm N}$ upon cooling for the $n$=3 and $n$=5 compounds that exhibit the AFM order at low temperatures, as shown in Figs.~\ref{fig:n5_PG} and \ref{fig:n3_PG}, respectively. 
In underdoped regions where the ground state is characterized by the coexistence of HTSC and AFM, low-lying magnetic excitations develop upon cooling toward the AFM order without a gap opening. 
In this context, unless an AFM order occurs in the underdoped region, it is natural for spin-singlet formation to develop above $T_{\rm c}$, leading to a pseudogap behavior in magnetic excitations.  Recently, Yamase {\it et al.} have theoretically pointed out that a {\it spin gap} is strongly suppressed near a tetracritical point for phases of the AFM, HTSC, AFM+HTSC, and normal states~\cite{Yamase2}. As a result, it is likely that the pseudogap, which emerges around the antinode region at the wave vectors (0,$\pi$) and ($\pi$,0), evolves to a real {\it AFM gap} in the AFM-HTSC mixed phase. One underlying issue is to experimentally address whether or not a pseudogap survives even in the paramagnetic and normal states in the AFM-HTSC mixed region.

\subsection{Momentum dependences of SC gap and pseudogap in HTSC}

\begin{figure}[tbp]
\centering
\includegraphics[width=7.5cm]{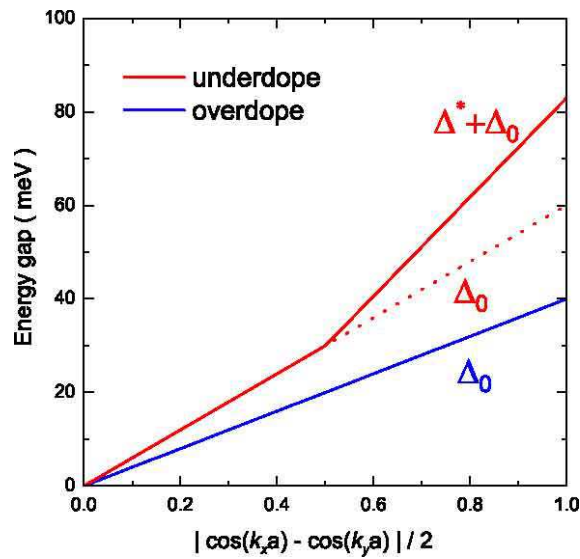}
\caption[]{\footnotesize (Color online) Illustration of momentum dependences of energy gap  $\Delta_0$ and pseudogap $\Delta^*$~\cite{Ideta}.[cited from ref.\cite{KitaokaIOP}]
}
\label{ARPES}
\end{figure}

As observed in ARPES experiments on multilayered cuprates~\cite{YChen,Ideta}, the momentum dependence of the gap magnitude in HTSC is schematically drawn in Fig.~\ref{ARPES}. Here, the gap for overdoped HTSC is almost a simple $d$ wave, $\Delta_0|\cos(k_xa)-\cos(k_ya)|/2$, as shown by a straight line.  On the other hand, the gap for the underdoped one deviates from the simple $d$-wave around the antinodes $\sim(0,\pi$) and ($\pi,0$). The gap size is characterized by two parameters, i.e., $\Delta_0$ around the node and $\Delta^*$ in the antinodal region, where $\Delta_0$ and $\Delta^*$ are defined by the linear extrapolation of the gap magnitude to the antinode [$\Delta_0|\cos(k_xa)-\cos(k_ya)|/2=\Delta_0$], as shown in the figure. The deviation of the gap anisotropy from the simple $d$-wave is known to be prominent in underdoped cuprates, which is called the {\it two-gap behavior}\cite{Tanaka2gap}. 

The ARPES on $n$=4:0234F($\sharp$4) with $T_{\rm c}$=55 K showed two Fermi sheets of IP and OP and revealed that their momentum dependences of the gap magnitude exhibit a {\it two-gap behavior} at IP and OP at 20 K~\cite{YChen}. When incorporating that the AFM order sets in at $T_{\rm N}$=80 K well above $T_{\rm c}$=55 K and that the AFM unit cell doubles a crystal one, $\Delta^*$ around the antinodes $\sim(0,\pi$) and ($\pi,0$) is assigned to an {\it AFM gap}. 
By contrast, $\Delta^*$, which used to be observed even in the  absence of an AFM order in the underdoped regime for the $n$=1 and 2 compounds, may be ascribed to a formation of a {\it spin gap} due to the development of a singlet resonating valence bond (RVB) state~\cite{Ogata}.

\subsection{Comments on underlying issues in underdoped cuprates}

The experimental results in Figs. \ref{fig:PhaseDiagram} and \ref{Mvsp} suggest that an AFM moment exists at Cu sites up to $p_c(M_{\rm AFM}$=$0)\le 0.14$ even in the $n$=1:LSCO and $n$=2:YBCO$_{6+x}$ compounds, but it cannot be observed as a static AFM order in the range of $p_c(n) < p \le p_c(M_{\rm AFM}$=$0)$ owing to strong 2D fluctuations since the interlayer magnetic coupling is too weak to stabilize a completely static AFM order with a 3D long-range character.
This instability of the ground state on the electronic states was corroborated by the facts that the application of a high magnetic field helps to bring about an AFM order with $M_{\rm AFM}\sim 0.1\mu_{\rm B}$ in the vortex state of LSCO with $p$=0.1~\cite{Lake} and YBCO$_{6.5}$\cite{Miller} and that the spin-glass phase existing in LSCO survives up to a hole density close to $p_c(M_{\rm AFM}$=$0)\sim 0.14$\cite{JulienSG}.  In addition,  the static AFM order was also observed when the charge-stripe order occurs at $x\sim$1/8 of LSCO~\cite{Tranquada}. The neutron scattering experiment suggested that an unusual AFM order with $M_{\rm AFM}\sim 0.1\mu_{\rm B}$ takes place in the underdoped YBCO$_{6.5}$ with $T_{\rm c}$=60 K, although it fluctuates on a nanosecond time scale~\cite{Sidis}. 
The phase diagrams of LSCO and YBCO$_{6+x}$ have been believed as the prototypes thus far; however, we note that the underlying issues in the underdoped region mentioned above should be reconsidered on the basis of the following facts: (i) an AFM moment intrinsically exists up to $p_c(M_{\rm AFM}$=$0) \le 0.14$ in the ground state, (ii) the interlayer magnetic coupling in  LSCO and YBCO$_{6+x}$ is not sufficient to stabilize a completely static AFM order and (iii) the chemical substitution for doping introduces inevitable disorders into the CuO$_2$ plane.

\section{Concluding Remarks}

Site-selective NMR studies on multilayered cuprates have unraveled the intrinsic phase diagram of homogeneously hole-doped CuO$_2$ plane as follows: 
\begin{enumerate}
\item[(i)]  The AFM {\it metallic} phase is robust and uniformly coexists with the HTSC phase up to the quantum critical hole density ($p_{c}(n)$), at which the AFM order collapses.
\item[(ii)]  $p_{c}(n)$ increases from 0.075 to 0.08 to 0.10 as the interlayer magnetic coupling becomes stronger in increasing from $n$=3 to 4 to 5, respectively. 
\item[(iii)]  The uniform coexistence of AFM and HTSC at a single CuO$_2$ plane is the universal phenomenon in the underdoped region for $p < p_c(n)$. 
\item[(iv)]  The maxima of $\Delta_{\rm SC}$ and $T_{\rm c}$ take place irrespective of $n$ at $p(T_{\rm c}^{\rm max})\sim$~0.16 just outside $p_c(M_{\rm AFM}$=$0)\sim 0.14$, at which the AFM moment inherent in the CuO$_2$ plane disappears in the ground state. 
\item[(v)] The pseudogap in magnetic excitations collapses in the AFM state for $p < p_c(n)$, where low-lying magnetic excitations develop upon cooling toward the AFM order. 
\item[(vi)]  The ground-state phase diagram of AFM and HTSC (see Fig.~\ref{Mvsp}) is in good agreement with the ground-state phase diagrams in terms of either the $t$-$J$ model~\cite{Chen,Giamarchi,Inaba,Anderson,Anderson1,Anderson2,Lee1,Himeda,Kotliar,Paramekanti1,Lee2,Yamase1,Yamase2,Paramekanti2,Shih1,Shih2,Ogata,Pathak} or the Hubbard model in a strong correlation regime~\cite{Senechal,Capone}.
\item[(vii)] The in-plane superexchange interaction $J_{\rm in}(p)$ for the $n$=4 and 5 compounds is larger than $J_{\rm in}(0)\sim$~1300 K for the infinite-layered AFM-Mott insulator Ca$_{0.85}$Sr$_{0.15}$CuO$_{2}$.
\end{enumerate}

In particular, we emphasize that the uniformly mixed phase of AFM and HTSC for $p < p_c(n)$ and the emergence of $d$-wave SC with the maximum of $T_{\rm c}$ just outside $p_c(M_{\rm AFM}$=$0)$ can be accounted for by the {\it Mott physics} based on the $t$-$J$ model. In the physics behind high-$T_{\rm c}$ phenomena, there is a very strong Coulomb repulsive interaction $U$~($ > 6$ eV), which  prohibits the double occupancy of an up-spin electron and a down-spin electron on the same site. When noting that $U$ is almost unchanged with doping, being almost the same as that in AFM-Mott insulators, a large $J_{\rm in}$ attracts electrons of opposite spins to be in neighboring sites, raising the $T_{\rm c}$ of cuprates up to as high as 160 K~\cite{Anderson1,Ogata,Anderson2}; there are no bosonic glues. 

\section*{Acknowledgement}

These works have been carried out in collaboration with M. Abe, Y. Yamaguchi, T. Sakaguchi, K. Itohara, S.-i. Tabata, S. Iwai, K. Matoba, Y. Araki, H. Kotegawa, Y. Tokunaga, K. Magishi, K. Ishida, G.-q. Zheng, and K. Asayama for NMR studies of multilayered cuprates. 
The samples were provided by P. M. Shirage, H. Kito, Y. Kodama, Y. Tanaka, H. Ihara (AIST), K. Tokiwa, and T. Watanabe (Tokyo University of Science). We thank M. Mori, T. Tohyama, S. Maekawa, H. Yamase, T. K. Lee, M. Ogata, and H. Fukuyama for their valuable discussions and comments. These works were supported by a Grant-in-Aid for Specially Promoted Research (20001004) and by the Global COE Program (Core Research and Engineering of Advanced Materials-Interdisciplinary Education Center for Materials Science) from the Ministry of Education, Culture, Sports, Science and Technology (MEXT), Japan.



\end{document}